\documentclass[]{emulateapj}

\begin{document}

\newcommand{\kms}{km~s$^{-1}$}
\newcommand{\msun}{$M_{\odot}$}
\newcommand{\rsun}{$R_{\odot}$}
\newcommand{\teff}{$T_{\rm eff}$}
\newcommand{\logg}{$\log{g}$}

\slugcomment{2016, ApJ, 818, 155}

\title{The ELM Survey. VII.  Orbital Properties of Low Mass White Dwarf Binaries\altaffilmark{*}}

\author{Warren R.\ Brown$^1$,
	A.\ Gianninas$^2$,
	Mukremin Kilic$^2$,
	Scott J.\ Kenyon$^1$,
	and Carlos Allende Prieto$^{3,4}$
	}

\affil{ $^1$Smithsonian Astrophysical Observatory, 60 Garden St, Cambridge, MA 02138 USA\\
	$^2$Homer L. Dodge Department of Physics and Astronomy, University of Oklahoma, 440 W. Brooks St., Norman, OK, 73019 USA\\
	$^3$Instituto de Astrof\'{\i}sica de Canarias, E-38205, La Laguna, Tenerife, Spain\\
	$^4$Departamento de Astrof\'{\i}sica, Universidad de La Laguna, E-38206 La Laguna, Tenerife, Spain
	}

\email{wbrown@cfa.harvard.edu, alexg@nhn.ou.edu, kilic@ou.edu}

\altaffiltext{*}{Based on observations obtained at the MMT Observatory, a joint 
facility of the Smithsonian Institution and the University of Arizona.}

\shorttitle{ Extremely Low Mass White Dwarfs. VII. }
\shortauthors{Brown et al.}

\begin{abstract}

	We present the discovery of 15 extremely low mass ($5 < \log{g} < 7$) white 
dwarf candidates, 9 of which are in ultra-compact double-degenerate binaries.  Our 
targeted ELM Survey sample now includes 76 binaries.  The sample has a lognormal 
distribution of orbital periods with a median period of 5.4 hr.  The velocity 
amplitudes imply that the binary companions have a normal distribution of mass with 
0.76 \msun\ mean and 0.25 \msun\ dispersion.  Thus extremely low mass white dwarfs 
are found in binaries with a typical mass ratio of 1:4.  Statistically speaking, 
95\% of the white dwarf binaries have a total mass below the Chandrasekhar mass and 
thus are not Type Ia supernova progenitors.  Yet half of the observed binaries will 
merge in less than 6 Gyr due to gravitational wave radiation; probable outcomes 
include single massive white dwarfs and stable mass transfer AM CVn binaries.

\end{abstract}


\section{INTRODUCTION}

	The ELM Survey is a targeted spectroscopic survey of extremely low mass 
(ELM) white dwarfs (WDs).  ELM WDs are interesting because the Universe is not old 
enough to form them through single-star evolution.  Instead, ELM WDs form out of 
binary common envelope evolution \citep{webbink84, iben90, marsh95}.  The result is 
that ELM WDs are the signposts of ultra-compact double-degenerate binaries, systems 
that are among the strongest known mHz gravitational wave sources \citep{brown11b, 
kilic15b}.


	We use the term ELM WD to describe a WD with surface gravity $5 \lesssim 
\log{g} \lesssim 7$ and effective temperature 8,$000~{\rm K} \lesssim T_{\rm eff} 
\lesssim 22,$000~K, an empirical definition inspired by the near complete absence of 
such objects in major spectroscopic WD catalogs \citep{eisenstein06, gianninas11, 
kleinman13, kepler15}.  We identify ELM WDs by their spectra:  hydrogen Balmer lines 
provide a sensitive measure of surface gravity in this effective temperature range.  
ELM WDs are distinct from helium-burning sdB stars that have $T_{\rm eff} > 
25$,000~K \citep{heber09} at these surface gravities.  ELM WDs are also distinct 
from hydrogen-burning main sequence stars that have $\log{g}<4.75$ at these 
temperatures \citep{kepler15b}.  Others have referred to similar objects as 
``low-mass WDs'' \citep{marsh95, moran97}, ``helium-core WD progenitors'' 
\citep{heber03, silvotti12}, and ``proto-WDs'' \citep{kaplan13}.  Theoretical WD 
evolutionary models indicate that there can indeed be some ambiguity about the 
evolutionary status of $\log{g} \sim 6$ hydrogen atmosphere objects:  thermonuclear 
hydrogen shell flashes, present in $>0.18$ \msun\ WD tracks, can cause WDs to loop 
around in \teff -$\log{g}$ space before they settle on the final cooling track 
\citep{sarna00, panei07, althaus13, istrate14}.  Observationally, ELM WDs in 
eclipsing binaries obey WD mass-radius relations \citep{steinfadt10, parsons11, 
vennes11, brown11b, kilic14b, hermes14, hallakoun15}.  Pulsating ELM WDs display 
pulsation frequencies that validate them as $\simeq$0.2 \msun\ degenerate objects 
\citep{hermes12b, hermes13a, hermes13b}.  We thus feel comfortable calling our 
$5<\log{g}<7$ objects ELM WDs, whether or not there is residual shell burning.  The 
recent discovery of a pulsating ELM WD in the highly relativistic binary PSR 
J1738+0333 will enable future, high-precision constraints on the internal structure 
and evolutionary state of an ELM WD \citep{kilic15}.

	Our approach to finding new ELM WDs is to select candidates by broadband 
color, and then to obtain spectroscopy to identify the nature of the objects.  
Previous ELM Survey papers have reported the discovery of 73 low mass WDs, 67 of 
which are in single-line spectroscopic binaries with orbital periods $P\le1$ day 
\citep{kilic10, kilic11a, kilic12a, brown10c, brown12a, brown13a, gianninas15}.  
The most spectacular object to-date is the $P=765$ sec detached eclipsing WD binary 
J0651, a gravitational wave source 10,000 times stronger than the Hulse-Taylor 
pulsar \citep{brown11b, hermes12c}.

	Here, we present 15 new ELM WD candidates, 9 of which are in short period 
binaries.  We will henceforth refer to any double-degenerate binary that contains an 
ELM WD as an ELM WD binary.  Six of the ELM WD candidates show no significant radial 
velocity variation.  While the non-variable objects have $\log{g}\simeq6$, the 
absence of orbital motion means that we cannot be certain whether or not they are 
ELM WDs. \citet{kepler15b} recently identified a population of ``sdA'' stars at 
comparable \logg\ and $T_{\rm eff} \sim 8$,000 K in the SDSS spectroscopic catalog.  
Given that many of our non-variable objects are clumped around 8,000 K, and that the 
number of our non-variable objects is in tension with the number of low inclination 
binaries expected in a random sample, at least some of the non-variable objects may 
be linked to this sdA stellar population.

	We focus our discussion on using the ELM WD binary sample to study the 
distribution of ELM WD orbital properties.  We begin by defining a clean sample of 
ELM WD binaries for detailed analysis.  We fit the observed distribution of orbital 
period and semi-amplitude, and then derive the distribution of companion mass $M_2$.  
The average companion is a 0.76 \msun\ WD, which implies an average binary mass 
ratio of 1:4 and a total binary mass below the Chandrasekhar mass.  The median 
gravitational merger time of our ELM WD binary sample is 6 Gyr; the likely merger 
outcomes are stable mass-transfer AM CVn systems and single massive 
hydrogen-deficient objects, such as extreme helium stars and R~CrB stars 
\citep{kilic10}.

\section{DATA}

\begin{figure}		
 \includegraphics[width=3.5in]{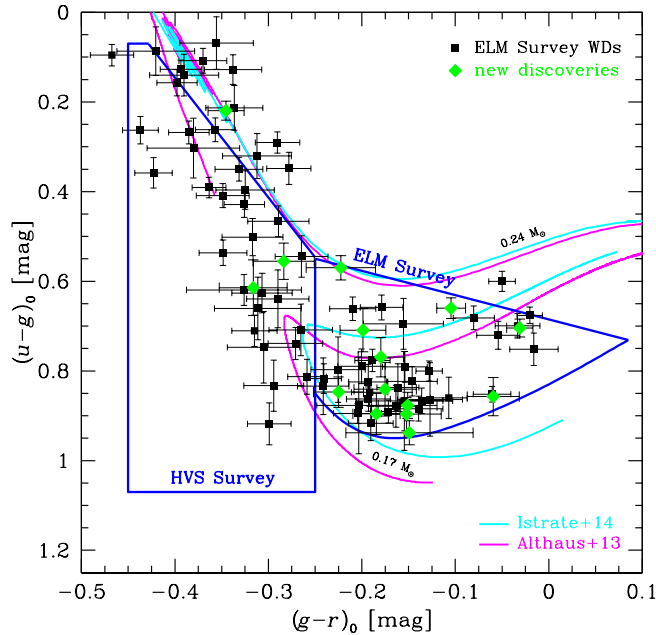}
 \caption{ \label{fig:ugr}
	Color-color plot of the ELM Survey.  Blue regions indicate our primary 
target selection regions.  New discoveries are marked as green diamonds; previously 
published WDs are marked as black squares.  Cyan and magenta lines show theoretical 
evolutionary tracks from \citet{istrate14} and \citet{althaus13}, respectively, for 
0.17 \msun\ and 0.24 \msun\ models discussed in Section 2.3.}
 \end{figure}

	We present observations of fifteen new ELM WD candidates.  Ten objects are 
selected by color for our targeted spectroscopic ELM Survey program as described in 
\citet{brown12a}.  Five objects come from follow-up spectroscopy of the completed 
Hypervelocity Star survey.  Hypervelocity Star survey targets are also selected by 
color, as described in \citet{brown12b, brown13a}.

	Figure \ref{fig:ugr} is a color-color plot showing ELM Survey objects in 
relation to our approximate target selection regions.  We unify the photometry for 
this paper using de-reddened, point spread function magnitudes from the Sloan 
Digital Sky Survey Data Release 12 \citep[SDSS,][]{alam15}.  We take reddening 
values from SDSS, and indicate de-reddened photometry and colors with a subscript 0.  
Objects fall outside our target selection regions for two reasons.  First, a handful 
of ELM WD candidates were selected from the SDSS spectroscopic catalog and thus were 
not drawn from the HVS Survey and ELM Survey target selection regions 
\citep{kilic11a, kilic12a}.  Second, SDSS photometry has changed since our original 
target selection:  of the five objects published here that come from the HVS Survey, 
three are now 1-2$\sigma$ outliers from the HVS Survey target selection region (see 
green diamonds in Figure \ref{fig:ugr}) presumably due to photometric 
recalibrations.  The SDSS colors for our objects thus have systematic errors 
comparable to their published statistical errors.

\begin{figure*}		
 \centerline{ \includegraphics[angle=270, width=6in]{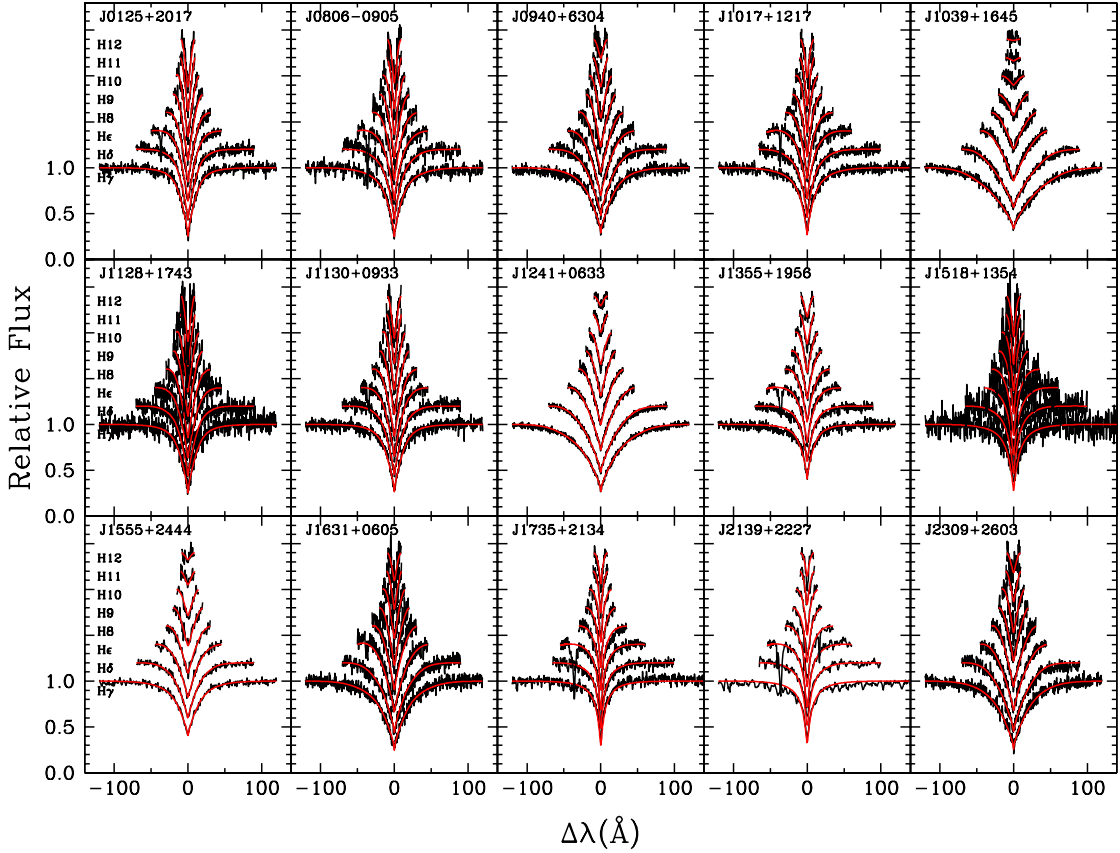} }
 \caption{ \label{fig:spec}
	Model fits ({smooth red lines}) overplotted on the composite observed 
spectra ({black lines}) for the 15 new low mass WDs. }
 \end{figure*}

\subsection{Spectroscopic Observations}

	We acquire spectra for the 15 ELM WD candidates using the Blue Channel 
spectrograph \citep{schmidt89} on the 6.5m MMT telescope.  We configured the Blue 
Channel spectrograph to obtain 3650 \AA\ -- 4500 \AA\ spectral coverage with 1.0 
\AA\ spectral resolution.  We acquire additional spectra for 5 objects using the 
KOSMOS spectrograph \citep{martini14} on the Kitt Peak National Observatory 4m 
Mayall telescope on program numbers 2014B-0119 and 2015A-0082.  We configured the 
KOSMOS spectrograph to obtain 3500 \AA\ -- 6200 \AA\ spectral coverage with 2.0 \AA\ 
spectral resolution.  We also acquire spectra for objects with $g<17$ mag using the 
FAST spectrograph \citep{fabricant98} on the Fred Lawrence Whipple Observatory 1.5m 
Tillinghast telescope.  We configured the FAST spectrograph to obtain 3500 \AA\ -- 
5500 \AA\ spectral coverage with 1.7 \AA\ spectral resolution.

	All spectra were paired with comparison lamp exposures for accurate 
wavelength calibration.  Flux calibration is performed with observations of blue 
spectrophotometric standards \citep{massey88}.  We measure radial velocities using 
the RVSAO cross-correlation program \citep{kurtz98} with high signal-to-noise 
templates.  Exposure times were chosen to yield a signal-to-noise ratio of 10 to 15 
per resolution element, resulting in typical radial velocity errors of 10 \kms\ to 
15 \kms\ for these WDs.

	Our observing strategy is to acquire a single spectrum per target, identify 
candidate ELM WDs from the initial spectra, then re-observe candidate ELM WDs in 
subsequent observing runs.  If a candidate showed velocity variability, we continue 
to observe the object until we constrain its orbital solution.  If a candidate shows 
no velocity variability but is confirmed to have $5<\log{g}<7$, we continue to 
observe it until we can rule out all $P<1$ day period aliases at high confidence.  
As a result, each object has around two dozen irregularly spaced observations 
obtained over a few year baseline.  The spectra published here were mostly acquired 
between 2013 January and 2015 October.  Figure \ref{fig:spec} presents the Balmer 
line profiles from the summed, rest-frame spectra for our 15 objects.  The 
individual radial velocities are provided in a data table presented in the 
Appendix.

\begin{figure}          
 \includegraphics[width=3.5in]{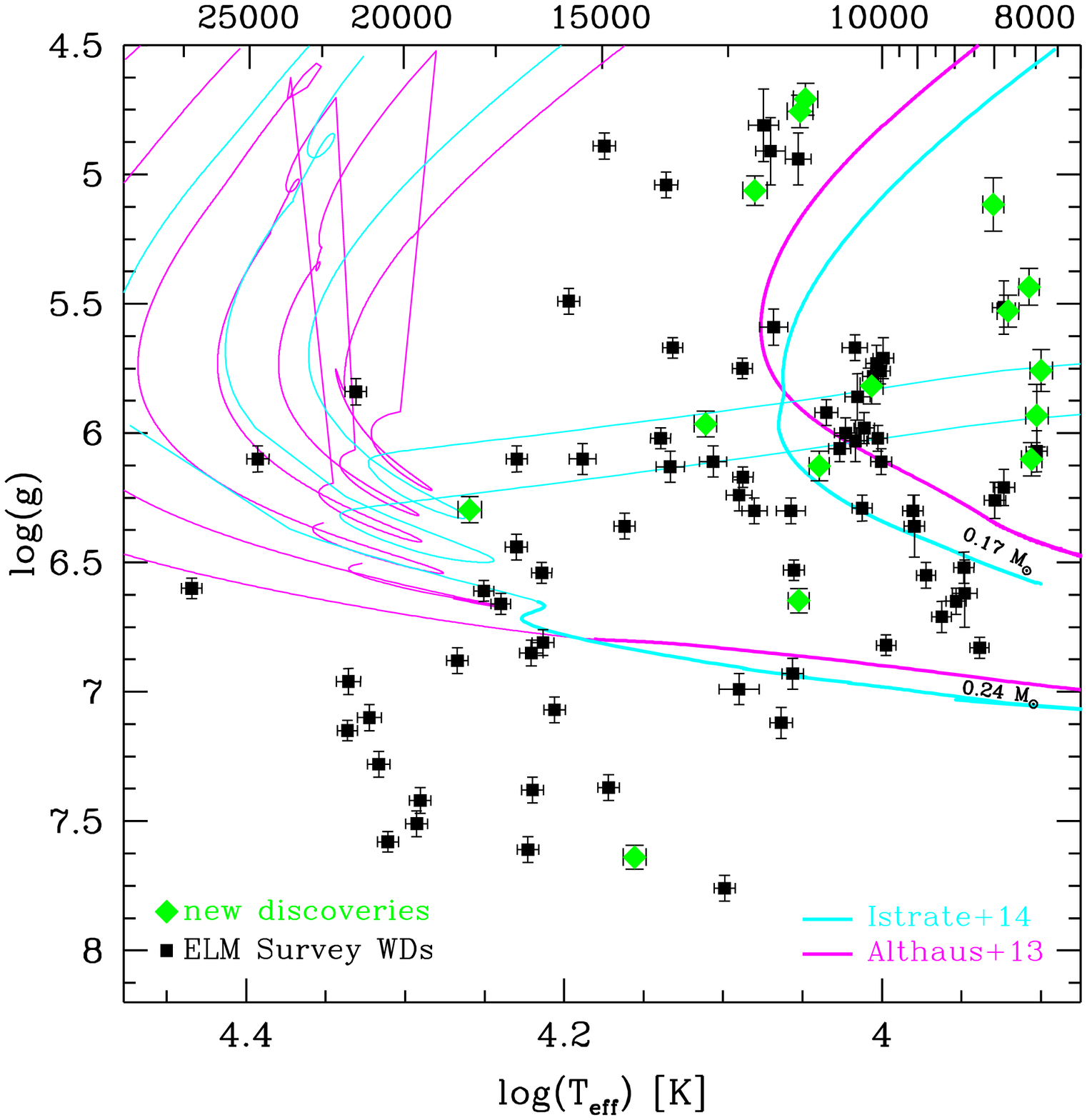}
 \caption{ \label{fig:teff}
	Surface gravity vs.\ effective temperature plot for the ELM Survey.  New 
discoveries are marked as green diamonds; previously published WDs are marked as 
black sqaures.  Cyan and magenta lines show evolutionary tracks from 
\citet{istrate14} and \citet{althaus13}, respectively, for 0.17 \msun\ and 0.24 
\msun\ models.  Short-lived parts of the tracks are indicated by thin lines (upper 
left corner); long-lived parts of the tracks are indicated by thick lines (right 
side).  Our color selection effectively samples 8,$000 < T_{\rm eff} < 22$,000~K.}
 \end{figure}

\begin{deluxetable*}{ccccccccc}	
\tabletypesize{\scriptsize}
\tablecolumns{9}
\tablewidth{0pt}
\tablecaption{WD Physical Parameters\label{tab:teff}}
\tablehead{
	\colhead{Object} &
	\colhead{R.A.} &
	\colhead{Decl.} &
	\colhead{$T_{\rm eff}$} &
	\colhead{$\log g$} &
	\colhead{Mass} &
	\colhead{$g_0$} &
	\colhead{$M_g$} &
	\colhead{$d_{helio}$} \\
   & (h:m:s) & (d:m:s) & (K) & (cm s$^{-2}$) & (\msun) & (mag) & (mag) & (kpc)
}
	\startdata
J0125$+$2017 &  1:25:16.758 &  20:17:44.63 & $11170 \pm 200$ & $4.709 \pm 0.062$ & $0.184 \pm 0.010$ & $17.153 \pm 0.021$ & $ 4.88 \pm 0.18$  & $2.855 \pm 0.238$ \\
J0806$-$0905 &  8:06:53.308 &  -9:05:11.51 & $ 8510 \pm 130$ & $5.116 \pm 0.104$ & $0.166 \pm 0.011$ & $17.681 \pm 0.006$ & $ 6.88 \pm 0.22$  & $1.454 \pm 0.145$ \\
J0940$+$6304 &  9:40:08.729 &  63:04:27.40 & $12910 \pm 210$ & $5.964 \pm 0.050$ & $0.180 \pm 0.010$ & $19.618 \pm 0.026$ & $ 7.69 \pm 0.13$  & $2.436 \pm 0.150$ \\
J1017$+$1217 & 10:17:07.109 &  12:17:57.42 & $ 8330 \pm 130$ & $5.528 \pm 0.062$ & $0.142 \pm 0.012$ & $17.479 \pm 0.020$ & $ 8.13 \pm 0.23$  & $0.745 \pm 0.083$ \\
J1039$+$1645 & 10:39:53.118 &  16:45:24.28 & $14310 \pm 240$ & $7.639 \pm 0.046$ & $0.458 \pm 0.018$ & $18.941 \pm 0.024$ & $10.71 \pm 0.08$  & $0.444 \pm 0.018$ \\
J1128$+$1743 & 11:28:23.334 &  17:43:54.58 & $11260 \pm 210$ & $4.756 \pm 0.063$ & $0.183 \pm 0.010$ & $19.434 \pm 0.026$ & $ 4.99 \pm 0.18$  & $7.774 \pm 0.656$ \\
J1130$+$0933 & 11:30:27.956 &   9:33:03.55 & $12020 \pm 210$ & $5.062 \pm 0.057$ & $0.179 \pm 0.010$ & $16.843 \pm 0.029$ & $ 5.62 \pm 0.16$  & $1.761 \pm 0.133$ \\
J1241$+$0633 & 12:41:24.291 &   6:33:51.02 & $11280 \pm 170$ & $6.648 \pm 0.047$ & $0.199 \pm 0.012$ & $17.722 \pm 0.021$ & $ 9.64 \pm 0.09$  & $0.413 \pm 0.018$ \\
J1355$+$1956 & 13:55:12.336 &  19:56:45.43 & $ 8050 \pm 120$ & $6.101 \pm 0.064$ & $0.156 \pm 0.010$ & $16.098 \pm 0.024$ & $ 9.76 \pm 0.17$  & $0.185 \pm 0.014$ \\
J1518$+$1354 & 15:18:02.566 &  13:54:31.96 & $ 8080 \pm 120$ & $5.435 \pm 0.071$ & $0.147 \pm 0.018$ & $18.988 \pm 0.019$ & $ 8.00 \pm 0.34$  & $1.594 \pm 0.246$ \\
J1555$+$2444 & 15:55:02.000 &  24:44:22.05 & $18170 \pm 310$ & $6.296 \pm 0.051$ & $0.190 \pm 0.012$ & $16.045 \pm 0.015$ & $ 7.82 \pm 0.12$  & $0.443 \pm 0.025$ \\
J1631$+$0605 & 16:31:23.675 &   6:05:33.82 & $10150 \pm 170$ & $5.818 \pm 0.069$ & $0.162 \pm 0.010$ & $19.002 \pm 0.019$ & $ 8.08 \pm 0.17$  & $1.533 \pm 0.118$ \\
J1735$+$2134 & 17:35:21.694 &  21:34:40.64 & $ 7940 \pm 130$ & $5.758 \pm 0.081$ & $0.142 \pm 0.010$ & $15.904 \pm 0.011$ & $ 9.06 \pm 0.21$  & $0.235 \pm 0.023$ \\
J2139$+$2227 & 21:39:07.415 &  22:27:08.87 & $ 7990 \pm 130$ & $5.932 \pm 0.121$ & $0.149 \pm 0.011$ & $15.600 \pm 0.011$ & $ 9.42 \pm 0.29$  & $0.174 \pm 0.023$ \\
J2309$+$2603 & 23:09:19.904 &  26:03:46.69 & $10950 \pm 160$ & $6.127 \pm 0.057$ & $0.176 \pm 0.010$ & $18.986 \pm 0.016$ & $ 8.54 \pm 0.14$  & $1.229 \pm 0.080$ \\
\enddata

\tablecomments{\teff\ and \logg\ values are corrected for 3D effects following \citet{tremblay15}.
 Mass and $M_g$ values are estimated with \citet{althaus13} models.}

\end{deluxetable*}

\subsection{Stellar Atmosphere Parameters}

	We derive the stellar atmosphere parameters for each WD by fitting the 
summed, rest-frame spectra to a grid of pure hydrogen atmosphere models as described 
in \citet{gianninas11, gianninas14b, gianninas15}.  In brief, the models cover 
4,$000~{\rm K} < T_{\rm eff} < 35$,000~K and $4.5 < \log{g} < 9.5$ and include the 
Stark broadening profiles from \citet{tremblay09}.  We also apply the 
\citet{tremblay13, tremblay15} three-dimensional stellar atmosphere model 
corrections to fix the so-called ``\logg\ problem.'' These corrections can lower the 
one-dimensional stellar atmosphere model parameters by up to 500 K in \teff\ and 0.3 
dex in \logg.  Our statistical uncertainties are typically $\pm180$ K in \teff\ and 
$\pm0.07$ dex in \logg.

	The best-fit models are over-plotted on each spectrum in Figure 
\ref{fig:spec}.  We mask the region around the Ca {\sc ii} K line when doing 
the Balmer line fits. Twelve of our new WDs have $5 < \log{g} < 7$ and are thus ELM 
WDs.  Two objects, J0125+2017 and J1128+1743, have anomalously low surface gravities 
$\log{g}\simeq4.7$ consistent with metal-poor, young main sequence stars,
however they both show short-period orbital motion and are probable ELM WDs 
(see Section 3.2).  The last object, J1039+1645, is a slightly more massive 
$\log{g}=7.64$ WD that also shows orbital motion.  Table \ref{tab:teff} presents the 
stellar atmosphere parameters for all 15 objects.

	Figure \ref{fig:teff} plots the distribution of \teff\ and \logg\ for the 
ELM Survey objects.  Our $(g-r)_0$ target selection results in an approximate 
temperature selection of 8,$000~{\rm K} < T_{\rm eff} < 22$,000~K.  The observed 
range of surface gravity reflects our choice to follow-up those objects with initial 
surface gravity estimates $5\lesssim \log{g} \lesssim 7$.  The clump of new objects 
around 8,000 K is from the non-variable objects.

\begin{figure}          
 \plotone{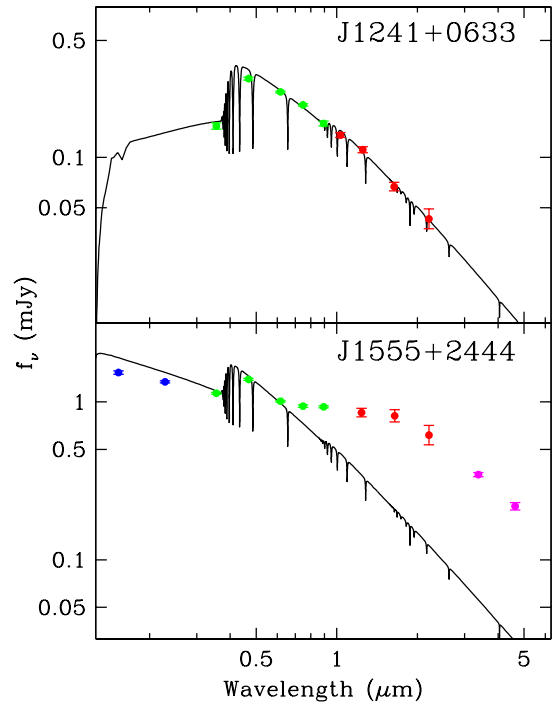}
 \caption{ \label{fig:sed}
	Comparison of the spectroscopic stellar atmosphere model (black line) 
to available photometry from GALEX (blue points), SDSS (green points), 2MASS or 
UKIDSS (red points), and WISE (magenta points).  J1241+0633 is a typical example, 
while J1555+2444 is a notable exception: a WD + M dwarf. }
 \end{figure}

\begin{deluxetable*}{ccccrcc}
\tabletypesize{\scriptsize}
\tablecolumns{7}
\tablewidth{0pt}
\tablecaption{Binary Orbital Parameters\label{tab:orbits}}
\tablehead{
\colhead{Object}&
\colhead{$N_{\rm obs}$}&
\colhead{$P$}&
\colhead{$k$}&
\colhead{$\gamma$}&
\colhead{$M_2$}&
\colhead{$\tau$}\\
  & & (days) & (km s$^{-1}$) & (km s$^{-1}$) & (\msun ) & (Gyr)
}
	\startdata
J0125$+$2017 & 16 & $0.88758 \pm 0.00004$ & $ 65.4 \pm  2.1$ & $  79.4 \pm  2.2$ & $>$0.14 & $<$938.3 \\
J0806$-$0905 & 16 &          \nodata      & $     < 28.5   $ & $ 109.4 \pm  3.7$ & \nodata &  \nodata \\
J0940$+$6304 & 20 & $0.48438 \pm 0.00001$ & $210.4 \pm  3.2$ & $  32.6 \pm  2.3$ & $>$0.73 & $<$51.27 \\
J1017$+$1217 & 14 &          \nodata      & $     < 30.2   $ & $  24.8 \pm  3.6$ & \nodata &  \nodata \\
J1039$+$1645 & 26 & $0.82470 \pm 0.02214$ & $ 83.4 \pm  4.0$ & $ -32.9 \pm  2.8$ & $>$0.31 & $<$186.2 \\
J1128$+$1743 & 21 & $2.16489 \pm 0.03889$ & $ 41.2 \pm  2.0$ & $ -24.5 \pm  1.4$ & $>$0.11 & $<$12350 \\
J1130$+$0933 & 16 & $1.55935 \pm 0.00014$ & $ 69.0 \pm  3.9$ & $-145.6 \pm  1.8$ & $>$0.19 & $<$3241  \\
J1241$+$0633 & 40 & $0.95912 \pm 0.00028$ & $138.2 \pm  4.8$ & $  90.0 \pm  3.3$ & $>$0.51 & $<$377.6 \\
J1355$+$1956 & 33 &          \nodata      & $     < 40.9   $ & $ -41.8 \pm  3.9$ & \nodata &  \nodata \\
J1518$+$1354 & 16 & $0.57655 \pm 0.00734$ & $112.7 \pm  4.6$ & $ -43.4 \pm  3.0$ & $>$0.23 & $<$237.6 \\
J1555$+$2444 & 20 &          \nodata      & $     < 23.0   $ & $   1.0 \pm  3.2$ & \nodata &  \nodata \\
J1631$+$0605 &  9 & $0.24776 \pm 0.00411$ & $215.4 \pm  3.4$ & $  -9.6 \pm  3.7$ & $>$0.47 & $<$13.15 \\
J1735$+$2134 & 20 &          \nodata      & $     < 31.6   $ & $ -16.3 \pm  3.4$ & \nodata &  \nodata \\
J2139$+$2227 & 37 &          \nodata      & $     < 22.0   $ & $   8.5 \pm  2.7$ & \nodata &  \nodata \\
J2309$+$2603 & 15 & $0.07653 \pm 0.00001$ & $405.8 \pm  3.5$ & $ -14.8 \pm  2.6$ & $>$0.79 & $<$0.359 \\
	\enddata
\end{deluxetable*}

\begin{figure*}			
 \plottwo{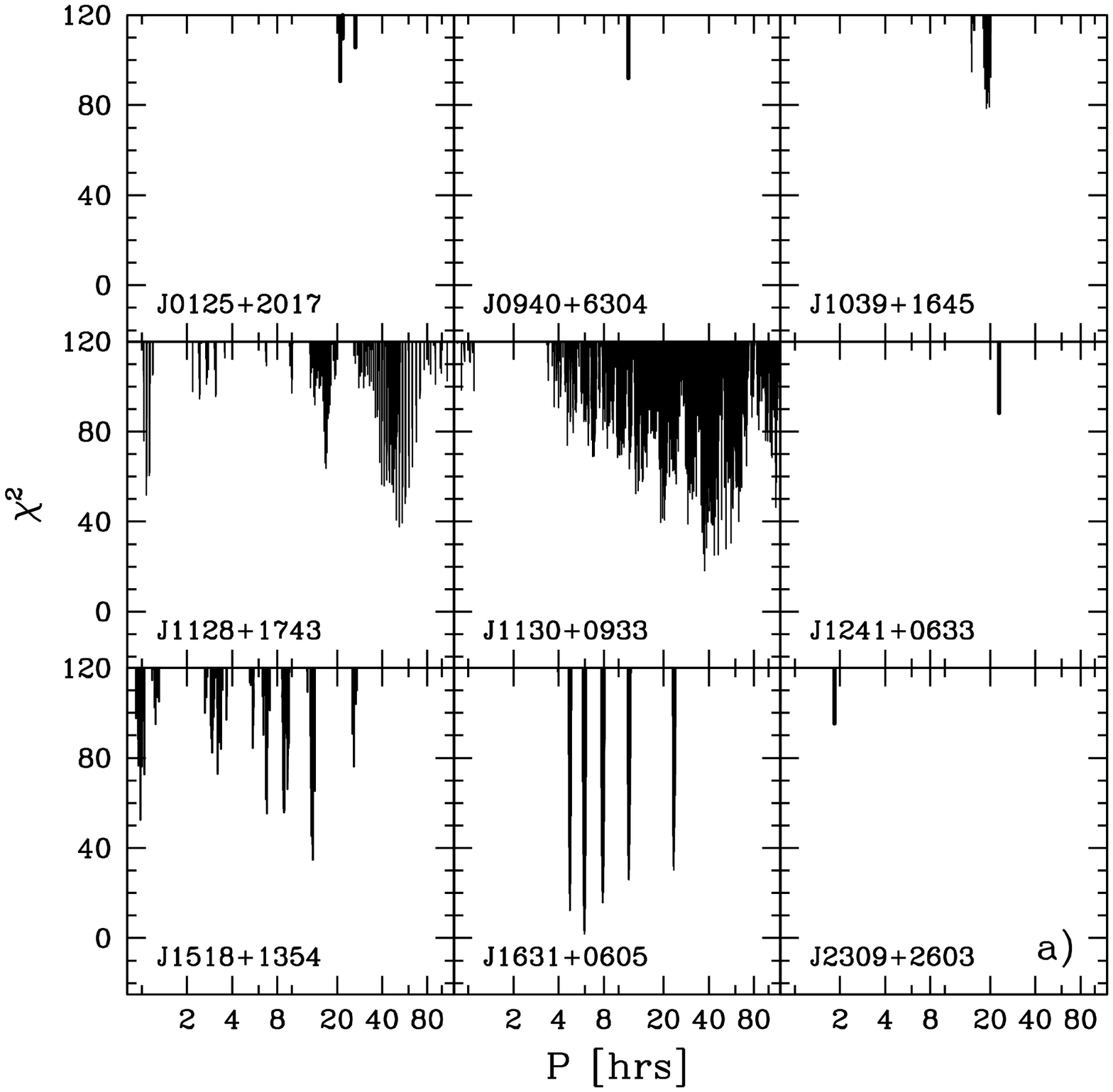}{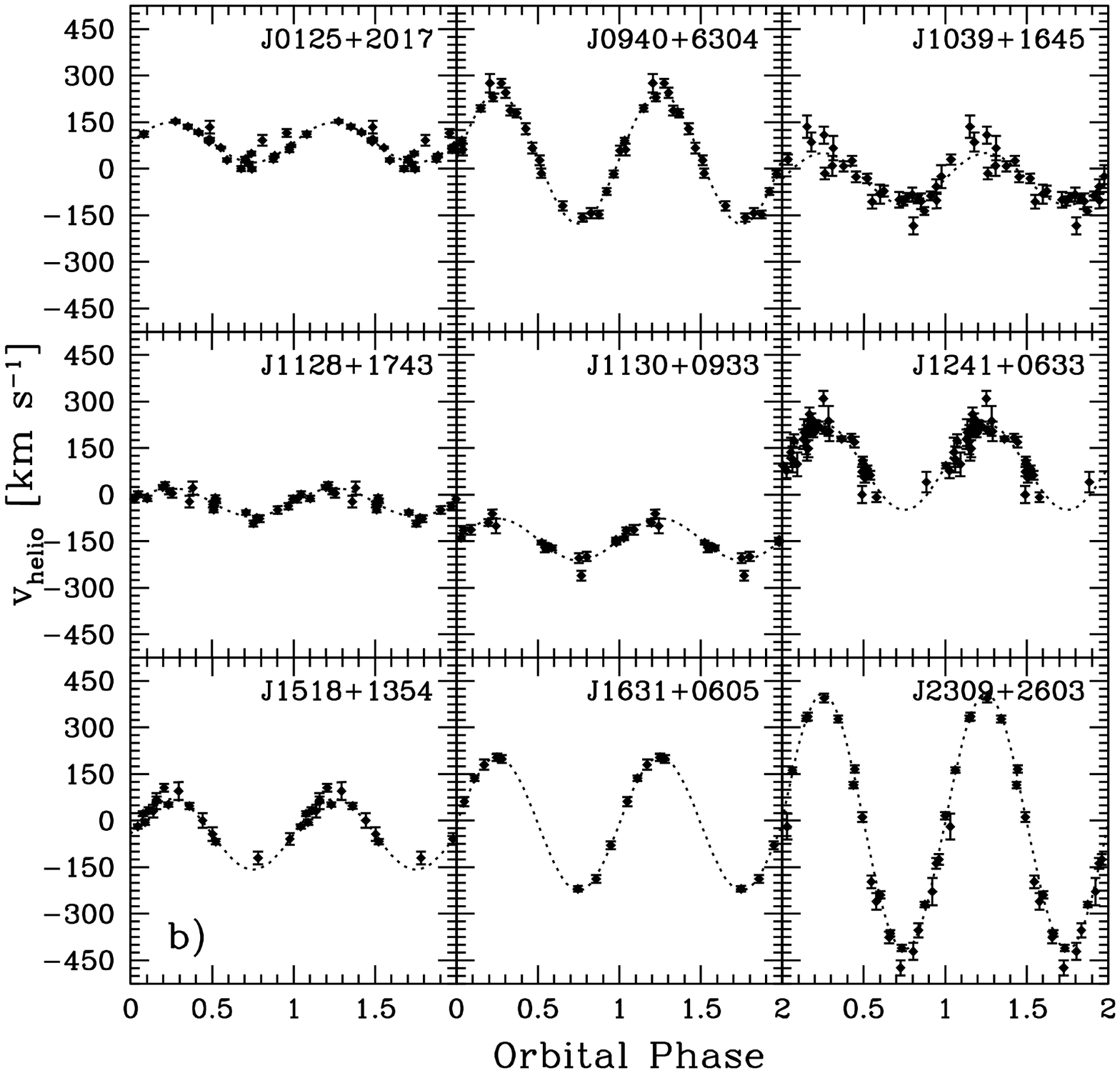}
 \caption{ \label{fig:pdm}
	a) Periodograms for the 9 WDs with significant orbital motion.  The best 
orbital periods have the smallest $\chi^2$.  Period aliases, with the exception 
of $P\simeq2$ day aliases in J1130+0933, are at least $\Delta \chi^2=15$ to 20 
larger than the minima and are likely insignificant. b) Observed velocities phased 
to the best fit orbits.}
 \end{figure*}

	We validate our stellar atmosphere parameters by comparing the best 
spectroscopic model for each object against photometry from GALEX \citep{martin05}, 
SDSS \citep{alam15}, 2MASS \citep{skrutskie06}, UKIDSS \citep{lawrence07}, and WISE 
\citep{wright10}.  We do not derive parameters from the spectral energy 
distributions because many of the fainter ELM WDs lack ultraviolet and infrared 
photometry.  Instead, we compare available photometry to the best spectroscopic 
fit as a consistency check.  We generally find excellent agreement between the 
models and the observed spectral energy distributions.  J1241+0633, for example, has 
a reduced $\chi^2$ of 1.15 (see Figure \ref{fig:sed}). A notable exception is 
J1555+2444, the first WD + M dwarf in the ELM Survey, which has a reduced 
$\chi^2$ of 181. WD + M dwarf binaries are common \citep[e.g.][]{rebassa12}, 
however our $ugri$ color selection is designed to exclude such systems.  In the case 
of J1555+2444, the M dwarf contributes significant flux in the near-infrared bands 
(see Figure \ref{fig:sed}).  We discuss this system further below.

\subsection{White Dwarf Parameters}

	We estimate WD mass and luminosity using the recent ELM WD evolutionary 
tracks of \citet{althaus13} and of \citet{istrate14}.  Both sets of tracks adopt 
neutron star companions for their evolutionary calculations, but an ELM WD's 
evolution should be unaffected by its companion once it detaches from the common 
envelope.  The choice of progenitor metallicity, on the other hand, can have a 
significant impact on the hydrogen envelope mass and resulting cooling times 
\citep{althaus15}.  Both sets of ELM WD tracks assume solar metallicity appropriate 
for disk objects; our ELM Survey sample contains both disk and halo objects.

	For reference, we plot fiducial 0.17 \msun\ and 0.24 \msun\ ELM WD tracks in 
Figures \ref{fig:ugr} and \ref{fig:teff}.  We draw \citet{althaus13} tracks in 
magenta and \citet{istrate14} tracks in cyan.  In Figure \ref{fig:ugr}, we plot 
estimated synthetic colors to compare with our color selection.  In Figure 
\ref{fig:teff}, we plot the published tracks, and use thin lines to indicate the 
short-lived parts of the tracks and thick lines to indicate the long-lived parts of 
the tracks.  Shell flashes, which generate loops and lead to faster evolution, are 
present in the 0.24 \msun\ tracks and absent in the 0.17 \msun\ tracks.

	We interpolate the evolutionary tracks to determine the mass and luminosity 
for each of our objects, however the interpolation is complicated by the fact that 
the tracks map phase space in a discrete and irregular way.  Our solution is to 
identify the two nearest tracks to an observed \teff\ and \logg\ value, and 
interpolate between those two tracks on the basis of \logg.  We then Monte Carlo our 
\teff\ and \logg\ with their uncertainties to derive the mass and luminosity 
uncertainties.  

	The evolutionary tracks of \citet{althaus13} and \citet{istrate14} yield 
very similar mass and luminosity estimates for our objects.  Consider the ELM Survey 
WDs in the mass range over which the two sets of tracks overlap:  $0.16 < M < 0.30$ 
\msun.  For these 66 objects, the two sets of tracks differ on average by 0.007 
\msun\ (systematic) $\pm0.012$ \msun\ (statistical) in mass and 0.044 mag 
(systematic) $\pm0.068$ mag (statistical) in absolute $g$-band magnitude $M_g$.  
\citet{istrate14} tracks yield slightly more massive and more luminous estimates, 
likely due to the lack of gravitational settling in the models resulting in mixed 
H/He atmospheres.  In any case, \citet{althaus13} tracks span a wider range of mass 
and so are more convenient to use with the entire ELM Survey sample.  We adopt 
\citet{althaus13} tracks for the following discussion, and add 0.01 \msun\ in 
quadrature to the formal mass uncertainty to account for the uncertainty in the 
choice of model.  Table \ref{tab:teff} presents the parameters, including the 
de-reddened apparent $g$-band magnitudes and distances, for all 15 WDs.

\subsection{Orbital Elements}

	Before determining orbital elements, we must first separate the radial 
velocity variable and non-variable objects.  We use the F-test to quantify how the 
variance of the radial velocities around a constant mean velocity compares to the 
variance around a best-fit orbital solution \citep[e.g.,][]{bevington92}.  The 
larger the F-test probability, the more likely the orbital solution is consistent 
with noise.  Six of the ELM WDs published here are significantly non-variable, with 
F-test probabilities $>0.01$ and typical 95\% confidence upper-limits on 
semi-amplitude of $k < 30$ \kms.  We discuss the non-variable ELM WDs further below.

	Nine WDs have significant velocity variability, with F-test probabilities 
$<0.002$.  We calculate orbital elements for these nine WDs in the same way as in 
prior ELM Survey papers.  In brief, we use the summed spectra as cross-correlation 
templates to maximize the velocity precision for each individual object, and then 
minimize $\chi^2$ for a circular orbit using the code of \citet{kenyon86}.  We 
search for orbital periods up to 5 days, which is the maximum time span of our 
individual observing runs and a period at which a 0.6 \msun\ WD companion (assuming 
$i=60\arcdeg$) produces a detectable $k=75$ \kms.  Figure \ref{fig:pdm} shows 
periodograms and phased radial velocity plots for the nine WDs.  We estimate the 
significance of the period aliases using $\chi^2$ values.  For normally distributed 
errors, a $\Delta \chi^2=13.3$ with respect to the minimum value corresponds to a 
99\% confidence interval for 4 degrees of freedom \citep{press92}.  On this 
basis, J1130+0933 has significant period aliases around 2 days, but we consider 
these aliases uninteresting given that they are longer than the best 1.6 day 
period.  For all other objects, the period aliases have $\Delta \chi^2=15$ to 20 
larger than the minimum and thus appear insignificant.  Table \ref{tab:orbits} 
presents the orbital elements.  The full ELM Survey sample of 88 objects is 
provided in a data table presented in the Appendix.

\subsection{ELM Survey Completeness}

	Over the magnitude range $15 < g_0 < 20$, we have obtained spectra for 99\% 
of the stars in the HVS Survey target selection region \citep{brown12a} and 80\% of 
the stars in the ELM Survey target selection region (Figure \ref{fig:ugr}).  In 
absolute numbers, 120 of 589 ELM Survey targets remain unobserved.  Almost all of 
the 120 unobserved targets are located between RA=21 hrs and RA=3 hrs due to 
telescope time allocation and weather.  Based on current detection rates, we 
estimate that 26\% of the remaining targets are likely DA WDs, of which one third 
(or about 10) are likely ELM WD binaries.

	Our multi-epoch follow-up spectroscopy is less complete.  We currently 
have 26 ELM WD candidates that require confirming observations, and another 17 
velocity variable ELM WDs that have poorly constrained orbital parameters.  Summing 
these numbers together with the unobserved targets suggests there may be another 
$\simeq$53 ELM WD binaries in our target selection regions.  Our published sample 
currently contains 76 binaries.  We therefore estimate that the published ELM Survey 
sample is approximately 60\% complete for ELM WD binaries.

\section{RESULTS}

	We discuss the 15 new ELM WD candidates in this section, starting with the 
1.8 hr orbital period binary system J2309+2603.  The other eight double 
degenerate binaries fill out the longer orbital period portion of our sample.  Our 
understanding of these systems is based on multi-epoch spectroscopy and broadband 
colors; we do not yet have time-series photometry for these systems.  We investigate 
the possible ELM WD + M dwarf system J1555+2444, and discuss the statistics of the 
non-velocity variable ELM WD candidates.

\subsection{J2309+2603}

	SDSS J230919.904+260346.69 (hereafter J2309+2603) is a 0.176 \msun\ ELM WD 
with a $1.8367\pm 0.0002$ hr orbital period and a $412.4\pm 2.7$ \kms\ 
semi-amplitude.  Given Kepler's 3rd Law, the minimum mass companion to this ELM WD 
is a 0.82 \msun\ object at an orbital separation of 0.76 \rsun.  If we adopt the 
mean inclination angle for a random stellar sample, $i = 60\arcdeg$, the companion 
is a 1.14 \msun\ object at an orbital separation of 0.83 \rsun.  There is no 
evidence for mass-transfer in this system.  Thus the ELM WD's most likely companion 
is another WD.

	There remains a 20\% likelihood, assuming a random distribution of 
inclinations, that the companion is a $1.4 < M_2 < 3.0$ \msun\ neutron star.  Given 
that this putative neutron star would have accreted material during the common 
envelope evolution of the ELM WD progenitor, J2309+2603 is possibly a millisecond 
pulsar binary system.  However no gamma ray source exists at this position in the 
{\it Fermi} Large Area Telescope source catalog \citep{acero15}.  A targeted {\it 
Chandra} search of other high-probability candidates in the ELM Survey has also 
found no neutron star companions \citep{kilic14b}.  We conclude that J2309+2603 is 
most likely a double WD binary with a total binary mass near the Chandrasekhar mass.

	J2309+2603 is losing energy and angular momentum to gravitational wave 
radiation.  The timescale for the binary to shrink and begin mass transfer via 
Roche-lobe overflow is given by the gravitational wave merger time
	\begin{equation} \label{eqn:gw}
\tau = 47925 \frac{(M_1 + M_2)^{1/3}}{M_1 M_2} P^{8/3} ~{\rm Myr}
        \end{equation} where the masses are in \msun\ and the period $P$ is in days 
\citep{kraft62}.  The merger time is 350 Myr for the minimum mass companion, and 
shorter if the companion is more massive.  

\subsection{Eight Other Double Degenerate Binaries}

	The other eight double degenerate binaries are longer orbital period, 
lower semi-amplitude systems (see Figure \ref{fig:pdm}).  The radial velocity 
constraints vary by object, and there is little to say about the best-measured 
system.  J0940+6304 has complete phase coverage and excellent constraints on period 
and semi-amplitude.

	J1241+0633 and J1518+1354 have excellent phase coverage and orbital period 
constraints, but increased semi-amplitude uncertainties due to a relative lack of 
observations near velocity minimum.  J1631+0605 is a $P=5.946\pm0.099$ hr binary 
with a well-constrained period and semi-amplitude -- observations were obtained at 
quadrature on back-to-back nights -- however incomplete phase coverage allows for 
the possibility of a modestly ($e\le0.3$) eccentric orbit.  Given that none of the 
other 75 binaries in the ELM Survey have significant eccentricity, nor do we expect 
much eccentricity from the common envelope origin of the ELM WD, we adopt the 
circular orbit solution.  J1039+1645, J1128+1743, and J1130+0933, by comparison, 
have excellent phase coverage but exhibit period aliases clumped around their 
best-fit periods.  Given that the observations rule out significantly different 
orbital periods for each object, we consider the 3\% uncertainty in period 
acceptable.  Systems with $P>16$ hr orbital periods have $>100$ Gyr merger times.  
The median likelihood for a neutron star companion in each of the eight 
longer-period binaries is 4\%.

	We note that J0125+2017 and J1128+1743 are abnormally low surface gravity 
objects, with $\log{g}\simeq4.7$, $T_{\rm eff} \simeq 11,200$ K, and masses around 
0.19 \msun\ if they are ELM WDs.  Their low $k\simeq50$ \kms\ semi-amplitudes imply 
that the unseen companions are comparable 0.2 \msun\ mass objects at 2.5 \rsun\ 
orbital separations (assuming $i = 60\arcdeg$).  Alternatively, it is possible these 
objects are very metal poor, young main sequence stars.  Padova tracks show that a 
1.6 \msun, $Z=0.0001$ star has similar \teff\ and \logg\ at $<100$ Myr ages 
\citep{bressan12}.  If J0125+2017 and J1128+1743 are very metal poor main sequence 
stars, their orbital motion would be due to 0.4 \msun\ mass companions at 4.5 \rsun\ 
orbital separations.  Very metal poor main sequence stars should not be forming near 
the Sun, but blue stragglers are possible \citep[e.g.][]{brown08b}.  However, 
observed blue stragglers have typical orbital periods of $\sim$1000 days 
\citep{geller12, geller15}.  Because J0125+2017 and J1128+1743 have orbital periods 
of 21 hr and 52 hr, respectively, we believe they are most likely ELM WD binaries.


\subsection{J1555+2444}

	J1555+2444 is a possible ELM WD + M dwarf system.  The M dwarf is classified 
as an 0.43 \msun\ M2 star \citep{rebassa10, schreiber10}.  Assuming that the M2 star 
dominates the $z$-band flux and has absolute magnitude $M_z=+8.0$ \citep{bressan12}, 
its distance is about 0.51 kpc. This is consistent with the 0.44 kpc and 0.49 kpc 
distance estimates to the hot \teff = 18,170 K ELM WD using \citet{althaus13} and 
\citet{istrate14} tracks, respectively.  Thus J1555+2444 could be a legitimate WD + 
M dwarf system.

	The ELM WD shows no orbital motion, however.  We constrain its orbital 
semi-amplitude to have a 95\% confidence upper limit of $k<23$ \kms\ on the basis of 
6 epochs of radial velocities obtained over 2 yrs.  This result is in tension with 
the work of \citet{gomezmoran11}, who find that the period distribution of WD + main 
sequence binaries peaks around $P=10.3$ hr.  The orbital period of J1555+2444 would 
have to be 3 months, assuming $i=60\arcdeg$, to be consistent with the observed 
semi-amplitude limit.  Alternatively, the orbital inclination would have to be 
$i<8\arcdeg$, assuming $P=10.3$ hr, to be consistent with the observed 
semi-amplitude limit, an inclination that has a 1\% likelihood in a random 
distribution.

	An ELM WD + M dwarf binary is also unlikely from a stellar evolutionary 
standpoint.  The standard ELM WD formation scenario requires double common-envelope 
evolution in which the companion star evolves first.  Other possibilities are that 
this is a triple system in which the M dwarf is an outer third, or else a chance 
super-position of an M dwarf and ELM WD.  J1555+2444 thus shares some similarities 
to J0935+4411, the 20 min orbital period double WD binary that appears to have an M 
dwarf along the line-of-sight \citep{kilic14}.  Additional observational constraints 
are needed to understand the nature of this object.

\begin{figure}          
 \includegraphics[width=3.5in]{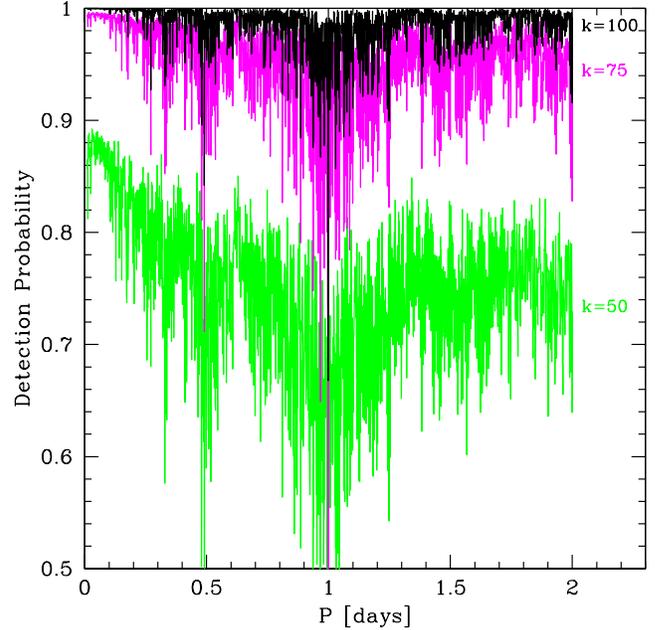}
 \caption{ \label{fig:pktest}
	Likelihood of detecting $k=50$ (green), 75 (magenta), and 100 (black) \kms\ 
orbital motion as a function of period $P$ given our set of observations for the six 
non-variable ELM WDs.}
 \end{figure}

\subsection{Non-Variable Objects:  ELM WDs or sdA Stars}

	Given that ELM WDs are the likely end product of compact binary evolution, 
the discovery of J1555+2444 and the five other non-variable objects is surprising.  
An obvious question is whether our observations have missed potential companions. We 
explore the sensitivity of our radial velocities to different possible orbital 
periods and semi-amplitudes following the procedure described by \citet{brown13a}: 
we Monte-Carlo the observed velocities with their errors, and count the number of 
10,000 orbital fits that have F-test probabilities $<0.01$ for orbital periods 
between 0.1 and 2 days.

	Figure \ref{fig:pktest} presents the averaged result for the six 
non-variable ELM WD candidates.  The lines in Figure \ref{fig:pktest} represent the 
probability of detecting $k=50$, 75, and 100 \kms\ orbits as a function of $P$ given 
our set of observations.  The objects have 6 or 7 epochs of time-series spectroscopy 
and a total of a couple dozen observations each.  Figure \ref{fig:pktest} shows 
that, while our ground-based observations are naturally less sensitive around 12 hr 
and 24 hr orbital periods, we are sensitive to almost any orbit with $k>75$ \kms.  
On average, our observations should detect 95\% of binary systems with $k=75$ \kms\ 
and 99\% of binary systems with $k=100$ \kms\ out to $P=2$ day orbital periods.  
Yet our observations show that the non-variable systems have a typical 95\% 
confidence upper limit of $k<30$ \kms.

	While we expect to find some binaries in face-on orientations, it is hard to 
explain all of the non-variable ELM WD candidates in this way.  Restricting 
ourselves to the 78 best ELM WD candidates in our Survey, those objects with 
$\log{g}<7.15$, we find 67 binaries and 11 non-variable objects.  The median 
semi-amplitude of the ELM WD binaries is $k=220$ \kms, a semi-amplitude that would 
appear $<75$ \kms\ for $i<20\arcdeg$ inclinations.  Assuming a random distribution 
of inclinations, only 5 of the 78 objects with $\log{g}<7.15$ should have 
$i<20\arcdeg$.  Yet 11 of the 78 objects are non-variable.  If we adopt the 
uncertainty as $\sqrt{N}$, the excess of non-variable objects appears significant at 
the 2.5-$\sigma$ level.  This excess suggests that the non-variable ELM 
WD candidates in our sample may not all be face-on binary systems.

	To avoid possible bias in relying on observed semi-amplitudes, we can 
instead assume a companion mass distribution $M_2$ and ask what fraction of binaries 
should fall below our detection threshold.  If we adopt the lognormal orbital period 
distribution (allowing $0<P<\infty$) and the normal $M_2$ distribution derived below 
(Section 4) and assume a random distribution of inclinations, 6.5 of 78 binaries 
should have $k<75$ \kms.  This number is in 2-$\sigma$ disagreement with the number 
of observed non-variable ELM WD candidates.  While not formally significant, it is 
suggestive that \citet{maxted00} find a similar excess of non-variable objects in 
their sample of $<0.5$ \msun\ low mass WDs.  Thus our non-variable objects are 
possibly binaries with long orbital periods, or they are possibly something else 
altogether.

	Notably, 8 of the 11 non-variable ELM WD candidates are found clumped 
together around $T_{\rm eff} \simeq 8$,000~K at the cool edge of our Survey. 
\citet{kepler15b} recently identified thousands of similar ``sdA'' stars with 
$T_{\rm eff} \sim 8$,000~K and $\log{g} \sim 6$ in the SDSS spectroscopic catalog.  
The non-variable ELM candidates are possibly linked to these sdA stars, however the 
sdA stars cannot be ELM WDs on the basis of their numbers.  According to the 
evolutionary tracks, an ELM WD spends about as much time with $7,500 < T_{\rm eff} < 
8,500$ K as it spends at 8,$500~{\rm K} < T_{\rm eff} < 22$,000 K.  Shell flashes 
can cause the time spent in these temperature ranges to differ by a factor of 2, but 
not by a factor of 10.  The models also show that 8,000 K ELM WDs are about 3 mag 
fainter in $M_g$ than 12,000 K ELM WDs.  Taken together, a magnitude-limited survey 
should observe fewer ELM WDs at 8,000 K than at higher temperatures.  Yet the sdA 
stars of \citet{kepler15b} are over 10 times more abundant than our higher-\teff\ 
ELM WD binaries in the same footprint of sky.  We conclude that the sdA stars are 
probably unrelated to ELM WDs.

	Our 8 non-variable objects with $T_{\rm eff} \simeq 8$,000~K are possibly 
related to the sdA stars.  If true, this would explain why we find so many 
non-variable objects around 8,000 K.  The three hotter non-variable objects would 
then be consistent at the 1-$\sigma$ level with the number of ELM WDs we would 
expect to find in face-on binaries in our color-selected survey.  Given the 
ambiguity about the nature of the non-variable objects, we focus the remainder of 
this paper on the ELM WD binaries.

\section{ANALYSIS AND DISCUSSION}

\subsection{Clean Sample}

	We begin by defining a clean sample of ELM WD binaries, and then use that 
sample to derive the orbital period and secondary mass distributions of the sample.
	We construct a clean sample of ELM WD binaries by first taking the 88 
published ELM Survey objects and excluding the 12 non-variable objects.  We further 
restrict the sample to those binaries with $k>75$ \kms, a semi-amplitude threshold 
above which our catalog should be 95\% complete (Figure \ref{fig:pktest}).  The ELM 
Survey color selection provides a built-in temperature selection, but the \logg\ 
selection is more ambiguous.  \citet{tremblay15} 3D corrections push some of the WD 
surface gravities below $\log{g}=5$.  We have also obtained observations for every 
WD near $\log{g}\simeq7$ during the course of our survey.  Since we would like to 
maximize our number statistics, we choose to restrict our sample to $4.85 < \log{g} 
< 7.15$, a range over which the ELM Survey observations should be reasonably 
complete.  Taken together, these cuts leave us with a clean sample of 62 ELM WD 
binaries.  We present the clean sample in Table \ref{tab:clean} sorted by orbital 
period.

\subsection{Orbital Period Distribution}

\begin{figure}          
 \plotone{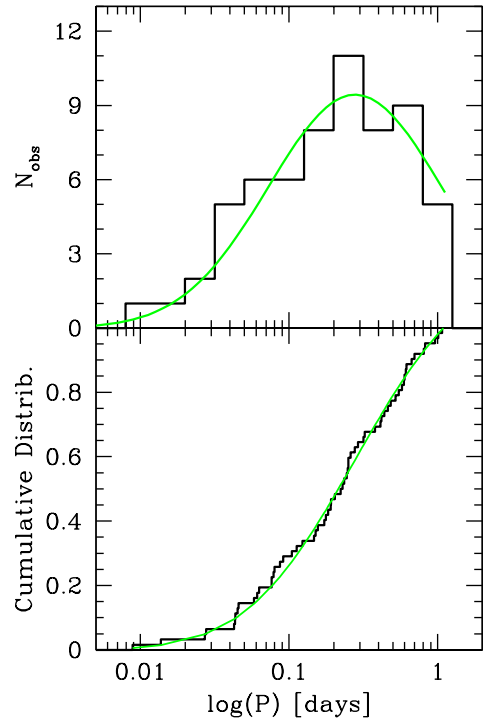}
 \caption{ \label{fig:pdist}
	Observed orbital period distribution of the clean sample of 62 ELM WD 
binaries (black lines) compared to a lognormal fit (green lines), plotted in a 
histogram (upper panel) and a cumulative distribution (lower panel).}
 \end{figure}

	Next, we consider the orbital period distribution of the clean sample, 
plotted in Figure \ref{fig:pdist}.  The median orbital period of the clean sample is 
0.226 days, or 5.4 hrs.

	We use the non-parametric Anderson-Darling test to evaluate the goodness of 
fit between a model distribution and the observations \citep{scholz87}.  The p-value 
of the Anderson-Darling test is the probability that the model and observations 
share a common distribution.  A p-value of zero rejects the null hypothesis that the 
model and observations share a common distribution.

	We fit different functional forms to the period distribution and find that 
the observations are best matched by a lognormal distribution,
	\begin{equation} f_P (P; \mu_P, \sigma_P^2) = \frac{1}{P\sqrt{2\pi}\sigma_P} 
{\rm exp}\left( -\frac{ (\ln{P} - \mu_P)^2 }{ 2\sigma_P^2 } \right)
	\end{equation} where P is the period, $\mu_P$ is the lognormal mean, and 
$\sigma_P$ is the standard deviation.  We restrict the fit to $P<1.1$ day to match 
the absence of orbital periods greater than $\simeq$1 day in the clean sample.  The 
green lines in Figure \ref{fig:pdist} show the best fit for lognormal mean $\mu_P = 
-1.28$ and standard deviation $\sigma_P=1.34$.  The Anderson-Darling p-value is  
0.999, indicating a strong consistency between the model and the observations.

\subsection{Secondary Mass Distribution}

	We now derive the (unseen) secondary mass distribution by considering the 
distribution of observables.  Formally, the observed orbital period $P$, 
semi-amplitude $k$, and derived ELM WD mass $M_1$ are related to the secondary mass 
$M_2$ and orbital inclination $i$ by the binary mass function,
	\begin{equation} \frac{P k^3}{2 \pi G} = \frac{(M_2 
\sin{i})^3}{(M_1+M_2)^2}. \label{eqn:mfn} \end{equation}
	Given our surface gravity constraints, the ELM WD mass distribution $M_1$ is 
narrow and centered around $\sim$0.2 \msun.  Thus most of the information about 
$M_2$ is encoded in the semi-amplitude distribution.  Inclinations of individual 
binaries are unknown, however the target selection is known:  the ELM WDs were 
targeted by color.  Thus we will assume that the inclination distribution is random 
in $\sin{i}$ above our detection threshold.

	Our computational approach is to Monte Carlo different trial distributions 
of $M_2$.  We draw inclination from a random $\sin{i}$ distribution and $P$ from the 
observed lognormal distribution.  We clip the simulations to match the observational 
$k>75$ \kms\ sample limit, a threshold that translates into an $i \gtrsim 20\arcdeg$ 
limit in inclination.  We then use the Anderson-Darling test to quantify how likely 
the simulated and observed distributions of semi-amplitude share a common 
distribution.

\begin{figure}          
 \includegraphics[width=3.5in]{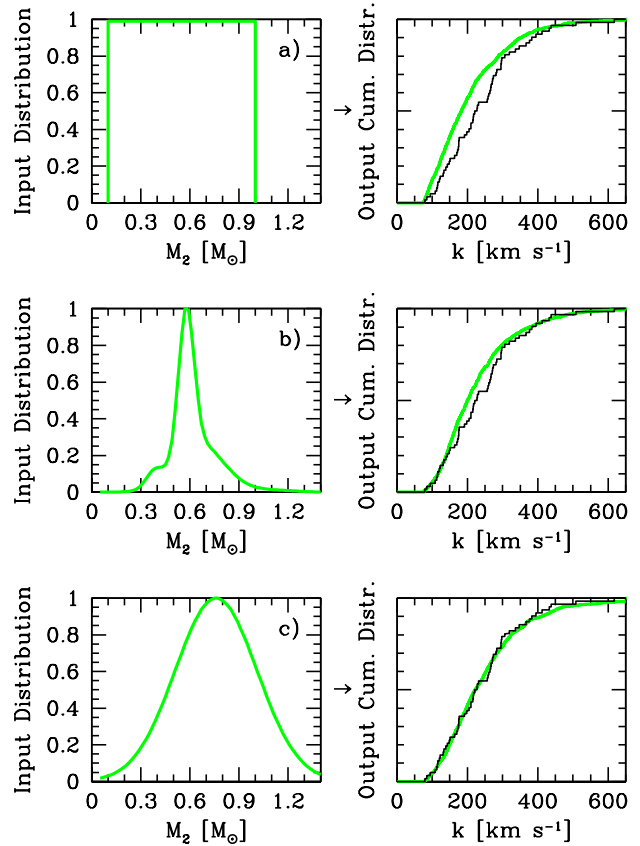}
 \caption{ \label{fig:m2}
	Trial companion mass $M_2$ distributions (lefthand column) and their 
simulated $k$ distributions (righthand column, in green) compared with the clean 
sample (in black).  a) Flat $M_2$ distribution observed in sdB binaries, $p=0.0012$.  
b) Observed SDSS WD mass distribution, $p=0.028$.  c) Normal $M_2$ distribution, 
$p=0.240$. }
 \end{figure}

	Figure \ref{fig:m2} presents three trial $M_2$ distributions and their 
corresponding simulated $k$ distributions compared to the observations.  At the top 
is the flat companion mass distribution observed in sdB binaries \citep{kupfer15}.  
The flat $M_2$ distribution produces a $k$ distribution that clearly differs from 
the observations ($p=0.0012$).  In the middle is the WD mass distribution observed 
by SDSS \citep{kepler07}. The SDSS WD mass distribution produces a $k$ distribution 
that agrees somewhat better with the observations ($p=0.028$).  Finally, we present 
what we find to be the best match:  a normal distribution with mean $\mu_{M2} = 
0.76$ \msun\ and standard deviation $\sigma_{M2} = 0.25$ \msun.  This normal 
distribution produces a $k$ distribution that agrees with the clean sample of ELM WD 
binaries with $p=0.240$.

	\citet{andrews14} derive essentially the same normal distribution using an 
earlier version of the ELM Survey data and a different Bayesian modeling approach.  
\citet{boffin15} also finds a similar result using an earlier version of the ELM 
Survey data and an inversion technique.  This agreement provides us with confidence 
in the normal distribution result.  Clearly, to fit the observed distribution of $k$ 
requires a large fraction of relatively massive WD companions.

\subsection{Implications}

	With the $M_2$ distribution in hand, we can go back and constrain the mass 
ratio, total mass, and merger time of the individual ELM WD binaries.  For some of 
the systems we have additional constraints on inclination:  two systems are 
eclipsing \citep{brown11b, kilic14b}, eight have ellipsoidal variations 
\citep{kilic11b, hermes12a, hermes14}, and many more have X-ray and/or radio 
observations that rule out massive neutron star companions \citep{kilic11a, 
kilic12a, kilic14b}.  Combining this information with our measurements of $P$, $k$, 
and $M_1$, we calculate the distributions of $M_1/M_2$, $M_{\rm tot}$, and $\tau$ 
for each system.  The results are presented in Table \ref{tab:clean}.  The columns 
list the median value of each distribution; the uncertainty comes from the 
0.1587 and 0.8413 percentile values, equivalent to 1-$\sigma$ uncertainties if the 
quantities are normally distributed.

	We find that the average total mass of the ELM WD binaries is $1.01\pm0.15$ 
\msun.  Statistically speaking, 95\% of the ELM WD binaries have a total mass below 
the Chandrasekhar mass and thus are not Type Ia supernova progenitors.  This is 
unsurprising given the mass of the ELM WDs.  However, their short orbital periods 
and relatively massive binary companions yield a median gravitational wave 
merger time of 5.8 Gyr for the sample.

	The outcome of the ELM WD binary merger process depends in part on the 
stability of mass transfer.  The ELM WD has the largest diameter of two objects in a 
double-degenerate binary and thus will become the donor star, the star that will 
fill its Roche lobe and will transfer matter to its companion when the binary 
evolves into contact.

	A clear implication of the $M_2$ distribution is that the binary components 
have fairly extreme mass ratios, on average $M_1$:$M_2 = 1$:$3.9 _{-0.8} ^{+1.2}$.  
For this mass ratio, theorists predict that helium mass transfer rates from the 
donor WD should be below the rate of stable helium burning on the accretor WD 
\citep{marsh04, kaplan12, kremer15}.  Thus mass transfer should proceed stably over 
billion year timescales, and the ELM WD binaries will become AM CVn systems 
\citep{warner95, solheim10}.  However, this statement ignores an ELM WD's hydrogen 
envelope.  If the initial hydrogen mass transfer rate is super-Eddington, a common 
envelope may form and cause the two WDs to quickly merge into a single massive WD 
\citep{webbink84, han99, dan11}.  Material ejected during shell flashes may also 
form a common envelope and result in a merger \citep{shen15}.

	Interestingly, a small peak around 1 \msun\ is seen in the mass distribution 
of all WDs within 20 pc \citep{giammichele12}.  The same peak is also seen in the 
mass distribution of WDs in the SDSS WD catalog \citep[e.g.][]{kepler07, kepler15}. 
\citet{rebassa15} argue that the $\simeq1$ \msun\ peak in the WD mass distribution 
is more substantial than suggested in these earlier surveys.  Perhaps part of the 1 
\msun\ peak in the WD mass distribution is related to the ELM WD binaries.  The 
merger of a He+CO WD should eliminate the hydrogen atmosphere and look like an 
extreme helium star or R~CrB star \citep{paczynski71}.  Whether the ELM WD binaries 
evolve into stable mass-transfer AM CVn systems or single massive WDs remains 
unclear, but it is fair to say that the binary mass ratio favors the AM CVn outcome.

\section{CONCLUSION}

	We present 15 new ELM WD candidates, 9 of which are double degenerate 
binaries.  This brings our targeted ELM Survey sample to 88 objects and 76 double 
degenerate binaries.  The 12 non-variable objects in the sample are puzzling, 
because they exceed the number of face-on binaries we expect in a randomly oriented 
sample.  Most of the non-variable objects have $T_{\rm eff} \simeq 8$,000~K, 
however, and so may be related to the sdA stellar population uncovered by 
\citet{kepler15b}; the remaining non-variable objects would then be consistent with 
the number of face-on ELM WD binaries we would expect in a sample of color-selected 
targets.

	We show that the ELM WD binaries have a lognormal distribution of orbital 
period with a median period of 5.4 hr.  This median period is in agreement with 
expectations of binary population synthesis models \citep{han98}. The distribution 
of companion mass is best described by a normal distribution with a mean of 0.76 
\msun.  Alternative distributions, such as the observed SDSS WD mass distribution, 
fit the observations very poorly.  The upshot is that the total mass of our ELM WD 
binaries is about 1 \msun, their mass ratio is about $M_1$:$M_2=1$:4, and their 
median gravitational merger time is about 6 Gyr.

	An open question is the outcome of ELM WD binary mergers.  Their total mass 
rules out Type Ia supernovae.  Their mass ratios argue for long-lived stable 
mass-transfer AM CVn binaries.  Yet a merger into a single massive WD remains a 
possibility.  Merger rates may provide a useful discriminant. Binary population 
synthesis models predict formation rates that differ by a factor of ten for AM CVn 
and R~CrB stars \citep[e.g.][]{zhang14, karakas15}.  Our early ELM WD binary merger 
rate \citep{brown11a} is a woeful under-estimate given the subsequent discovery of 
four ELM WD binaries with $<25$ Myr merger times \citep{brown11b, kilic11b, 
kilic11c, kilic14}.  In the next paper, we will divide our growing and increasingly 
complete ELM WD binary sample into disk and halo objects, derive the space 
densities, and then estimate merger rates over the entire Milky Way.

\acknowledgments

	We thank Charlotte Wood for her contribution to the spectral energy 
distributions.  We thank E.\ Martin, A.\ Milone, and S. Gotilla for their assistance 
with observations obtained at the MMT Observatory, P.\ Canton for his assistance 
with observations obtained at Kitt Peak National Observatory, and P.\ Berlind and 
M.\ Calkins for their assistance with observations obtained at the Fred Lawrence 
Whipple Observatory.  This project makes use of data products from SDSS and SDSS-II, 
funding for which was provided by the Alfred P.\ Sloan Foundation, the Participating 
Institutions, the National Science Foundation, the U.S.\ Department of Energy, the 
National Aeronautics and Space Administration, the Japanese Monbukagakusho, the Max 
Planck Society, and the Higher Education Funding Council for England.  This research 
makes use the SAO/NASA Astrophysics Data System Bibliographic Service.  This work 
was supported in part by the Smithsonian Institution.  MK and AG gratefully 
acknowledge the support of the NSF and NASA under grants AST-1312678 and NNX14AF65G, 
respectively.

{\it Facilities:} \facility{MMT (Blue Channel Spectrograph), FLWO:1.5m (FAST), 
Mayall (KOSMOS)}

	~


\begin{thebibliography}{91}
\expandafter\ifx\csname natexlab\endcsname\relax\def\natexlab#1{#1}\fi

\bibitem[{{Acero} {et~al.}(2015){Acero}, {Ackermann}, {Ajello},
  {et~al.}}]{acero15}
{Acero}, F., {Ackermann}, M., {Ajello}, M., {et~al.} 2015, \apjs, 218, 23

\bibitem[{{Alam} {et~al.}(2015){Alam}, {Albareti}, {Allende Prieto},
  {et~al.}}]{alam15}
{Alam}, S., {Albareti}, F.~D., {Allende Prieto}, C., {et~al.} 2015, \apjs, 219,
  12

\bibitem[{{Althaus} {et~al.}(2015){Althaus}, {Camisassa}, {Miller Bertolami},
  {C{\'o}rsico}, \& {Garc{\'{\i}}a-Berro}}]{althaus15}
{Althaus}, L.~G., {Camisassa}, M.~E., {Miller Bertolami}, M.~M., {C{\'o}rsico},
  A.~H., \& {Garc{\'{\i}}a-Berro}, E. 2015, \aap, 576, A9

\bibitem[{{Althaus} {et~al.}(2013){Althaus}, {Miller Bertolami}, \&
  {C{\'o}rsico}}]{althaus13}
{Althaus}, L.~G., {Miller Bertolami}, M.~M., \& {C{\'o}rsico}, A.~H. 2013,
  \aap, 557, A19

\bibitem[{{Andrews} {et~al.}(2014){Andrews}, {Price-Whelan}, \&
  {Ag{\"u}eros}}]{andrews14}
{Andrews}, J.~J., {Price-Whelan}, A.~M., \& {Ag{\"u}eros}, M.~A. 2014, ApJ Letters,
  797, L32

\bibitem[{{Bevington} \& {Robinson}(1992)}]{bevington92}
{Bevington}, P.~R. \& {Robinson}, D.~K. 1992, {Data reduction and error
  analysis for the physical sciences} (2nd ed.; New York: McGraw-Hill)

\bibitem[{{Boffin}(2015)}]{boffin15}
{Boffin}, H.~M.~J. 2015, \aap, 575, L13

\bibitem[{{Bressan} {et~al.}(2012){Bressan}, {Marigo}, {Girardi},
  {et~al.}}]{bressan12}
{Bressan}, A., {Marigo}, P., {Girardi}, L., {et~al.} 2012, \mnras, 427, 127

\bibitem[{{Brown} {et~al.}(2008){Brown}, {Beers}, {Wilhelm},
  {et~al.}}]{brown08b}
{Brown}, W.~R., {Beers}, T.~C., {Wilhelm}, R., {et~al.} 2008, \aj, 135, 564

\bibitem[{{Brown} {et~al.}(2012{\natexlab{a}}){Brown}, {Geller}, \&
  {Kenyon}}]{brown12b}
{Brown}, W.~R., {Geller}, M.~J., \& {Kenyon}, S.~J. 2012{\natexlab{a}}, \apj,
  751, 55

\bibitem[{{Brown} {et~al.}(2013){Brown}, {Kilic}, {Allende Prieto},
  {Gianninas}, \& {Kenyon}}]{brown13a}
{Brown}, W.~R., {Kilic}, M., {Allende Prieto}, C., {Gianninas}, A., \&
  {Kenyon}, S.~J. 2013, \apj, 769, 66

\bibitem[{{Brown} {et~al.}(2010){Brown}, {Kilic}, {Allende Prieto}, \&
  {Kenyon}}]{brown10c}
{Brown}, W.~R., {Kilic}, M., {Allende Prieto}, C., \& {Kenyon}, S.~J. 2010,
  \apj, 723, 1072

\bibitem[{{Brown} {et~al.}(2011{\natexlab{a}}){Brown}, {Kilic}, {Allende
  Prieto}, \& {Kenyon}}]{brown11a}
---. 2011{\natexlab{a}}, \mnras, 411, L31

\bibitem[{{Brown} {et~al.}(2012{\natexlab{b}}){Brown}, {Kilic}, {Allende
  Prieto}, \& {Kenyon}}]{brown12a}
---. 2012{\natexlab{b}}, \apj, 744, 142

\bibitem[{{Brown} {et~al.}(2011{\natexlab{b}}){Brown}, {Kilic}, {Hermes},
  {et~al.}}]{brown11b}
{Brown}, W.~R., {Kilic}, M., {Hermes}, J.~J., {et~al.} 2011{\natexlab{b}},
  ApJ Letters, 737, L23

\bibitem[{{Dan} {et~al.}(2011){Dan}, {Rosswog}, {Guillochon}, \&
  {Ramirez-Ruiz}}]{dan11}
{Dan}, M., {Rosswog}, S., {Guillochon}, J., \& {Ramirez-Ruiz}, E. 2011, \apj,
  737, 89

\bibitem[{{Eisenstein} {et~al.}(2006){Eisenstein}, {Liebert}, {Harris},
  {et~al.}}]{eisenstein06}
{Eisenstein}, D.~J., {Liebert}, J., {Harris}, H.~C., {et~al.} 2006, \apjs, 167,
  40

\bibitem[{{Fabricant} {et~al.}(1998){Fabricant}, {Cheimets}, {Caldwell}, \&
  {Geary}}]{fabricant98}
{Fabricant}, D., {Cheimets}, P., {Caldwell}, N., \& {Geary}, J. 1998, \pasp,
  110, 79

\bibitem[{{Geller} {et~al.}(2015){Geller}, {Latham}, \& {Mathieu}}]{geller15}
{Geller}, A.~M., {Latham}, D.~W., \& {Mathieu}, R.~D. 2015, \aj, 150, 97

\bibitem[{{Geller} \& {Mathieu}(2012)}]{geller12}
{Geller}, A.~M. \& {Mathieu}, R.~D. 2012, \aj, 144, 54

\bibitem[{{Giammichele} {et~al.}(2012){Giammichele}, {Bergeron}, \&
  {Dufour}}]{giammichele12}
{Giammichele}, N., {Bergeron}, P., \& {Dufour}, P. 2012, \apjs, 199, 29

\bibitem[{{Gianninas} {et~al.}(2011){Gianninas}, {Bergeron}, \&
  {Ruiz}}]{gianninas11}
{Gianninas}, A., {Bergeron}, P., \& {Ruiz}, M.~T. 2011, \apj, 743, 138

\bibitem[{{Gianninas} {et~al.}(2014){Gianninas}, {Dufour}, {Kilic},
  {et~al.}}]{gianninas14b}
{Gianninas}, A., {Dufour}, P., {Kilic}, M., {et~al.} 2014, \apj, 794, 35

\bibitem[{{Gianninas} {et~al.}(2015){Gianninas}, {Kilic}, {Brown}, {Canton}, \&
  {Kenyon}}]{gianninas15}
{Gianninas}, A., {Kilic}, M., {Brown}, W.~R., {Canton}, P., \& {Kenyon}, S.~J.
  2015, \apj, 812, 167

\bibitem[{{Hallakoun} {et~al.}(2015){Hallakoun}, {Maoz}, {Kilic},
  {et~al.}}]{hallakoun15}
{Hallakoun}, N., {Maoz}, D., {Kilic}, M., {et~al.} 2015, \mnras, submitted

\bibitem[{{Han}(1998)}]{han98}
{Han}, Z. 1998, \mnras, 296, 1019

\bibitem[{{Han} \& {Webbink}(1999)}]{han99}
{Han}, Z. \& {Webbink}, R.~F. 1999, \aap, 349, L17

\bibitem[{{Heber}(2009)}]{heber09}
{Heber}, U. 2009, \araa, 47, 211

\bibitem[{{Heber} {et~al.}(2003){Heber}, {Edelmann}, {Lisker}, \&
  {Napiwotzki}}]{heber03}
{Heber}, U., {Edelmann}, H., {Lisker}, T., \& {Napiwotzki}, R. 2003, \aap, 411,
  L477

\bibitem[{{Hermes} {et~al.}(2014){Hermes}, {Brown}, {Kilic},
  {et~al.}}]{hermes14}
{Hermes}, J.~J., {Brown}, W.~R., {Kilic}, M., {et~al.} 2014, \apj, 792, 39

\bibitem[{{Hermes} {et~al.}(2012{\natexlab{a}}){Hermes}, {Kilic}, {Brown},
  {Montgomery}, \& {Winget}}]{hermes12a}
{Hermes}, J.~J., {Kilic}, M., {Brown}, W.~R., {Montgomery}, M.~H., \& {Winget},
  D.~E. 2012{\natexlab{a}}, \apj, 749, 42

\bibitem[{{Hermes} {et~al.}(2012{\natexlab{b}}){Hermes}, {Kilic}, {Brown},
  {et~al.}}]{hermes12c}
{Hermes}, J.~J., {Kilic}, M., {Brown}, W.~R., {et~al.} 2012{\natexlab{b}},
  ApJ Letters, 757, L21

\bibitem[{{Hermes} {et~al.}(2013{\natexlab{a}}){Hermes}, {Montgomery},
  {Gianninas}, {et~al.}}]{hermes13b}
{Hermes}, J.~J., {Montgomery}, M.~H., {Gianninas}, A., {et~al.}
  2013{\natexlab{a}}, \mnras, 436, 3573

\bibitem[{{Hermes} {et~al.}(2012{\natexlab{c}}){Hermes}, {Montgomery},
  {Winget}, {et~al.}}]{hermes12b}
{Hermes}, J.~J., {Montgomery}, M.~H., {Winget}, D.~E., {et~al.}
  2012{\natexlab{c}}, ApJ Letters, 750, L28

\bibitem[{{Hermes} {et~al.}(2013{\natexlab{b}}){Hermes}, {Montgomery},
  {Winget}, {et~al.}}]{hermes13a}
---. 2013{\natexlab{b}}, \apj, 765, 102

\bibitem[{{Iben}(1990)}]{iben90}
{Iben}, Jr., I. 1990, \apj, 353, 215

\bibitem[{{Istrate} {et~al.}(2014){Istrate}, {Tauris}, {Langer}, \&
  {Antoniadis}}]{istrate14}
{Istrate}, A.~G., {Tauris}, T.~M., {Langer}, N., \& {Antoniadis}, J. 2014,
  \aap, 571, L3

\bibitem[{{Kaplan} {et~al.}(2013){Kaplan}, {Bhalerao}, {van Kerkwijk},
  {et~al.}}]{kaplan13}
{Kaplan}, D.~L., {Bhalerao}, V.~B., {van Kerkwijk}, M.~H., {et~al.} 2013, \apj,
  765, 158

\bibitem[{{Kaplan} {et~al.}(2012){Kaplan}, {Bildsten}, \&
  {Steinfadt}}]{kaplan12}
{Kaplan}, D.~L., {Bildsten}, L., \& {Steinfadt}, J.~D.~R. 2012, \apj, 758, 64

\bibitem[{{Karakas} {et~al.}(2015){Karakas}, {Ruiter}, \& {Hampel}}]{karakas15}
{Karakas}, A.~I., {Ruiter}, A.~J., \& {Hampel}, M. 2015, \apj, 809, 184

\bibitem[{{Kenyon} \& {Garcia}(1986)}]{kenyon86}
{Kenyon}, S.~J. \& {Garcia}, M.~R. 1986, \aj, 91, 125

\bibitem[{{Kepler} {et~al.}(2007){Kepler}, {Kleinman}, {Nitta},
  {et~al.}}]{kepler07}
{Kepler}, S.~O., {Kleinman}, S.~J., {Nitta}, A., {et~al.} 2007, \mnras, 375,
  1315

\bibitem[{{Kepler} {et~al.}(2015{\natexlab{a}}){Kepler}, {Pelisoli}, {Koester},
  {et~al.}}]{kepler15b}
{Kepler}, S.~O., {Pelisoli}, I., {Koester}, D., {et~al.} 2015{\natexlab{a}},
  \mnras, submitted

\bibitem[{{Kepler} {et~al.}(2015{\natexlab{b}}){Kepler}, {Pelisoli}, {Koester},
  {et~al.}}]{kepler15}
---. 2015{\natexlab{b}}, \mnras, 446, 4078

\bibitem[{{Kilic} {et~al.}(2010){Kilic}, {Brown}, {Allende Prieto}, {Kenyon},
  \& {Panei}}]{kilic10}
{Kilic}, M., {Brown}, W.~R., {Allende Prieto}, C., {Kenyon}, S.~J., \& {Panei},
  J.~A. 2010, \apj, 716, 122

\bibitem[{{Kilic} {et~al.}(2011{\natexlab{a}}){Kilic}, {Brown}, {Allende
  Prieto}, {et~al.}}]{kilic11a}
{Kilic}, M., {Brown}, W.~R., {Allende Prieto}, C., {et~al.} 2011{\natexlab{a}},
  \apj, 727, 3

\bibitem[{{Kilic} {et~al.}(2012){Kilic}, {Brown}, {Allende Prieto},
  {et~al.}}]{kilic12a}
---. 2012, \apj, 751, 141

\bibitem[{{Kilic} {et~al.}(2014{\natexlab{a}}){Kilic}, {Brown}, {Gianninas},
  {et~al.}}]{kilic14}
{Kilic}, M., {Brown}, W.~R., {Gianninas}, A., {et~al.} 2014{\natexlab{a}},
  \mnras, 444, L1

\bibitem[{{Kilic} {et~al.}(2015{\natexlab{a}}){Kilic}, {Brown}, {Hermes}, \&
  {Gianninas}}]{kilic15b}
{Kilic}, M., {Brown}, W.~R., {Hermes}, J.~J., \& {Gianninas}, A.
  2015{\natexlab{a}}, Astrophysics and Space Science Proceedings, 40, 167

\bibitem[{{Kilic} {et~al.}(2011{\natexlab{b}}){Kilic}, {Brown}, {Hermes},
  {et~al.}}]{kilic11c}
{Kilic}, M., {Brown}, W.~R., {Hermes}, J.~J., {et~al.} 2011{\natexlab{b}},
  \mnras, 418, L157

\bibitem[{{Kilic} {et~al.}(2011{\natexlab{c}}){Kilic}, {Brown}, {Kenyon},
  {et~al.}}]{kilic11b}
{Kilic}, M., {Brown}, W.~R., {Kenyon}, S.~J., {et~al.} 2011{\natexlab{c}},
  \mnras, 413, L101

\bibitem[{{Kilic} {et~al.}(2015{\natexlab{b}}){Kilic}, {Hermes}, {Gianninas},
  \& {Brown}}]{kilic15}
{Kilic}, M., {Hermes}, J.~J., {Gianninas}, A., \& {Brown}, W.~R.
  2015{\natexlab{b}}, \mnras, 446, L26

\bibitem[{{Kilic} {et~al.}(2014{\natexlab{b}}){Kilic}, {Hermes}, {Gianninas},
  {et~al.}}]{kilic14b}
{Kilic}, M., {Hermes}, J.~J., {Gianninas}, A., {et~al.} 2014{\natexlab{b}},
  \mnras, 438, L26

\bibitem[{{Kleinman} {et~al.}(2013){Kleinman}, {Kepler}, {Koester},
  {et~al.}}]{kleinman13}
{Kleinman}, S.~J., {Kepler}, S.~O., {Koester}, D., {et~al.} 2013, \apjs, 204, 5

\bibitem[{{Kraft} {et~al.}(1962){Kraft}, {Mathews}, \& {Greenstein}}]{kraft62}
{Kraft}, R.~P., {Mathews}, J., \& {Greenstein}, J.~L. 1962, \apj, 136, 312

\bibitem[{{Kremer} {et~al.}(2015){Kremer}, {Sepinsky}, \&
  {Kalogera}}]{kremer15}
{Kremer}, K., {Sepinsky}, J., \& {Kalogera}, V. 2015, \apj, 806, 76

\bibitem[{{Kupfer} {et~al.}(2015){Kupfer}, {Geier}, {Heber},
  {et~al.}}]{kupfer15}
{Kupfer}, T., {Geier}, S., {Heber}, U., {et~al.} 2015, \aap, 576, A44

\bibitem[{{Kurtz} \& {Mink}(1998)}]{kurtz98}
{Kurtz}, M.~J. \& {Mink}, D.~J. 1998, \pasp, 110, 934

\bibitem[{{Lawrence} {et~al.}(2007){Lawrence}, {Warren}, {Almaini},
  {et~al.}}]{lawrence07}
{Lawrence}, A., {Warren}, S.~J., {Almaini}, O., {et~al.} 2007, \mnras, 379,
  1599

\bibitem[{{Marsh} {et~al.}(1995){Marsh}, {Dhillon}, \& {Duck}}]{marsh95}
{Marsh}, T.~R., {Dhillon}, V.~S., \& {Duck}, S.~R. 1995, \mnras, 275, 828

\bibitem[{{Marsh} {et~al.}(2004){Marsh}, {Nelemans}, \& {Steeghs}}]{marsh04}
{Marsh}, T.~R., {Nelemans}, G., \& {Steeghs}, D. 2004, \mnras, 350, 113

\bibitem[{{Martin} {et~al.}(2005)}]{martin05}
{Martin}, D.~C. {et~al.} 2005, ApJ Letters, 619, L1

\bibitem[{{Martini} {et~al.}(2014){Martini}, {Elias}, {Points},
  {et~al.}}]{martini14}
{Martini}, P., {Elias}, J., {Points}, S., {et~al.} 2014, in Proc. SPIE 9147,
  Ground-based and Airborne Instrumentation for Astronomy V, ed. {Ramsay, S.~K
  and McLean, I.~S. and Takami, H.} (Montreal, Canada: SPIE), 91470Z

\bibitem[{{Massey} {et~al.}(1988){Massey}, {Strobel}, {Barnes}, \&
  {Anderson}}]{massey88}
{Massey}, P., {Strobel}, K., {Barnes}, J.~V., \& {Anderson}, E. 1988, \apj,
  328, 315

\bibitem[{{Maxted} {et~al.}(2000){Maxted}, {Marsh}, \& {Moran}}]{maxted00}
{Maxted}, P.~F.~L., {Marsh}, T.~R., \& {Moran}, C.~K.~J. 2000, \mnras, 319, 305

\bibitem[{{Moran} {et~al.}(1997){Moran}, {Marsh}, \& {Bragaglia}}]{moran97}
{Moran}, C., {Marsh}, T.~R., \& {Bragaglia}, A. 1997, \mnras, 288, 538

\bibitem[{{Nebot G{\'o}mez-Mor{\'a}n} {et~al.}(2011){Nebot
  G{\'o}mez-Mor{\'a}n}, {G{\"a}nsicke}, {Schreiber}, {et~al.}}]{gomezmoran11}
{Nebot G{\'o}mez-Mor{\'a}n}, A., {G{\"a}nsicke}, B.~T., {Schreiber}, M.~R.,
  {et~al.} 2011, \aap, 536, A43

\bibitem[{{Paczy{\'n}ski}(1971)}]{paczynski71}
{Paczy{\'n}ski}, B. 1971, Acta Astron., 21, 1

\bibitem[{{Panei} {et~al.}(2007){Panei}, {Althaus}, {Chen}, \& {Han}}]{panei07}
{Panei}, J.~A., {Althaus}, L.~G., {Chen}, X., \& {Han}, Z. 2007, \mnras, 382,
  779

\bibitem[{{Parsons} {et~al.}(2011){Parsons}, {Marsh}, {G{\"a}nsicke}, {Drake},
  \& {Koester}}]{parsons11}
{Parsons}, S.~G., {Marsh}, T.~R., {G{\"a}nsicke}, B.~T., {Drake}, A.~J., \&
  {Koester}, D. 2011, ApJ Letters, 735, L30

\bibitem[{{Press} {et~al.}(1992){Press}, {Teukolsky}, {Vetterling}, \&
  {Flannery}}]{press92}
{Press}, W.~H., {Teukolsky}, S.~A., {Vetterling}, W.~T., \& {Flannery}, B.~P.
  1992, {Numerical recipes in C. The art of scientific computing} (Cambridge:
  University Press, 2nd ed.)

\bibitem[{{Rebassa-Mansergas} {et~al.}(2010){Rebassa-Mansergas},
  {G{\"a}nsicke}, {Schreiber}, {Koester}, \&
  {Rodr{\'{\i}}guez-Gil}}]{rebassa10}
{Rebassa-Mansergas}, A., {G{\"a}nsicke}, B.~T., {Schreiber}, M.~R., {Koester},
  D., \& {Rodr{\'{\i}}guez-Gil}, P. 2010, \mnras, 402, 620

\bibitem[{{Rebassa-Mansergas} {et~al.}(2012){Rebassa-Mansergas}, {Nebot
  G{\'o}mez-Mor{\'a}n}, {Schreiber}, {et~al.}}]{rebassa12}
{Rebassa-Mansergas}, A., {Nebot G{\'o}mez-Mor{\'a}n}, A., {Schreiber}, M.~R.,
  {et~al.} 2012, \mnras, 419, 806

\bibitem[{{Rebassa-Mansergas} {et~al.}(2015){Rebassa-Mansergas}, {Rybicka},
  {Liu}, {Han}, \& {Garc{\'{\i}}a-Berro}}]{rebassa15}
{Rebassa-Mansergas}, A., {Rybicka}, M., {Liu}, X.-W., {Han}, Z., \&
  {Garc{\'{\i}}a-Berro}, E. 2015, \mnras, 452, 1637

\bibitem[{{Sarna} {et~al.}(2000){Sarna}, {Ergma}, \& {Ger{\v s}kevit{\v
  s}-Antipova}}]{sarna00}
{Sarna}, M.~J., {Ergma}, E., \& {Ger{\v s}kevit{\v s}-Antipova}, J. 2000,
  \mnras, 316, 84

\bibitem[{{Schmidt} {et~al.}(1989){Schmidt}, {Weymann}, \& {Foltz}}]{schmidt89}
{Schmidt}, G.~D., {Weymann}, R.~J., \& {Foltz}, C.~B. 1989, \pasp, 101, 713

\bibitem[{{Scholz} \& {Stephens}(1987)}]{scholz87}
{Scholz}, F.~W. \& {Stephens}, M.~A. 1987, Journal of the American Statistical
  Association, 82, 918

\bibitem[{{Schreiber} {et~al.}(2010){Schreiber}, {G{\"a}nsicke},
  {Rebassa-Mansergas}, {et~al.}}]{schreiber10}
{Schreiber}, M.~R., {G{\"a}nsicke}, B.~T., {Rebassa-Mansergas}, A., {et~al.}
  2010, \aap, 513, L7

\bibitem[{{Shen}(2015)}]{shen15}
{Shen}, K.~J. 2015, ApJ Letters, 805, L6

\bibitem[{{Silvotti} {et~al.}(2012){Silvotti}, {{\O}stensen}, {Bloemen},
  {et~al.}}]{silvotti12}
{Silvotti}, R., {{\O}stensen}, R.~H., {Bloemen}, S., {et~al.} 2012, \mnras,
  424, 1752

\bibitem[{{Skrutskie} {et~al.}(2006){Skrutskie}, {Cutri}, {Stiening},
  {et~al.}}]{skrutskie06}
{Skrutskie}, M.~F., {Cutri}, R.~M., {Stiening}, R., {et~al.} 2006, \aj, 131,
  1163

\bibitem[{{Solheim}(2010)}]{solheim10}
{Solheim}, J. 2010, \pasp, 122, 1133

\bibitem[{{Steinfadt} {et~al.}(2010){Steinfadt}, {Kaplan}, {Shporer},
  {Bildsten}, \& {Howell}}]{steinfadt10}
{Steinfadt}, J.~D.~R., {Kaplan}, D.~L., {Shporer}, A., {Bildsten}, L., \&
  {Howell}, S.~B. 2010, ApJ Letters, 716, L146

\bibitem[{{Tremblay} \& {Bergeron}(2009)}]{tremblay09}
{Tremblay}, P.-E. \& {Bergeron}, P. 2009, \apj, 696, 1755

\bibitem[{{Tremblay} {et~al.}(2015){Tremblay}, {Gianninas}, {Kilic},
  {et~al.}}]{tremblay15}
{Tremblay}, P.-E., {Gianninas}, A., {Kilic}, M., {et~al.} 2015, \apj, 809, 148

\bibitem[{{Tremblay} {et~al.}(2013){Tremblay}, {Ludwig}, {Steffen}, \&
  {Freytag}}]{tremblay13}
{Tremblay}, P.-E., {Ludwig}, H.-G., {Steffen}, M., \& {Freytag}, B. 2013, \aap,
  559, A104

\bibitem[{{Vennes} {et~al.}(2011){Vennes}, {Thorstensen}, {Kawka},
  {et~al.}}]{vennes11}
{Vennes}, S., {Thorstensen}, J.~R., {Kawka}, A., {et~al.} 2011, ApJ Letters, 737, L16

\bibitem[{{Warner}(1995)}]{warner95}
{Warner}, B. 1995, \apss, 225, 249

\bibitem[{{Webbink}(1984)}]{webbink84}
{Webbink}, R.~F. 1984, \apj, 277, 355

\bibitem[{{Wright} {et~al.}(2010){Wright}, {Eisenhardt}, {Mainzer},
  {et~al.}}]{wright10}
{Wright}, E.~L., {Eisenhardt}, P.~R.~M., {Mainzer}, A.~K., {et~al.} 2010, \aj,
  140, 1868

\bibitem[{{Zhang} {et~al.}(2014){Zhang}, {Jeffery}, {Chen}, \& {Han}}]{zhang14}
{Zhang}, X., {Jeffery}, C.~S., {Chen}, X., \& {Han}, Z. 2014, \mnras, 445, 660

\end{thebibliography}


\begin{deluxetable*}{ccccccc}
\tablecolumns{7}
\tablewidth{0pt}
\tablecaption{Clean Sample of ELM WD Binaries\label{tab:clean}}
\tablehead{
\colhead{Object}&
\colhead{$P$}&
\colhead{$k$}&
\colhead{$M_1$}&
\colhead{$M_1/M_2$}&
\colhead{$M_{\rm tot}$}&
\colhead{$\tau$}\\
\colhead{} & \colhead{(days)} & \colhead{(km s$^{-1}$)} & \colhead{(\msun)} &\colhead{} & \colhead{(\msun)} & \colhead{(Gyr)}
}
\startdata
0651$+$2844  &  $0.00886 \pm 0.00001$  &  $616.9 \pm 5.0$  &  $0.247 \pm 0.015$ & $ 0.50 _{-0.01} ^{+0.01}$  &  $ 0.74 _{-0.01} ^{+0.01}$  &  $  0.001 _{-0.0001}^{+0.0001}$ \\
0935$+$4411  &  $0.01375 \pm 0.00051$  &  $198.5 \pm 3.2$  &  $0.313 \pm 0.019$ & $ 0.42 _{-0.10} ^{+0.20}$  &  $ 1.07 _{-0.24} ^{+0.25}$  &  $  0.002 _{-0.0004}^{+0.0008}$ \\
0106$-$1000  &  $0.02715 \pm 0.00002$  &  $395.2 \pm 3.6$  &  $0.189 \pm 0.011$ & $ 0.33 _{-0.09} ^{+0.05}$  &  $ 0.75 _{-0.07} ^{+0.21}$  &  $  0.027 _{-0.006} ^{+0.003}$ \\
1630$+$4233  &  $0.02766 \pm 0.00004$  &  $295.9 \pm 4.9$  &  $0.298 \pm 0.019$ & $ 0.39 _{-0.10} ^{+0.17}$  &  $ 1.06 _{-0.23} ^{+0.24}$  &  $  0.015 _{-0.003} ^{+0.005}$ \\
1053$+$5200  &  $0.04256 \pm 0.00002$  &  $264.0 \pm 2.0$  &  $0.204 \pm 0.012$ & $ 0.27 _{-0.06} ^{+0.12}$  &  $ 0.97 _{-0.23} ^{+0.23}$  &  $  0.068 _{-0.012} ^{+0.021}$ \\
0056$-$0611  &  $0.04338 \pm 0.00002$  &  $376.9 \pm 2.4$  &  $0.180 \pm 0.010$ & $ 0.22 _{-0.03} ^{+0.03}$  &  $ 1.00 _{-0.10} ^{+0.13}$  &  $  0.076 _{-0.008} ^{+0.008}$ \\
1056$+$6536  &  $0.04351 \pm 0.00103$  &  $267.5 \pm 7.4$  &  $0.334 \pm 0.016$ & $ 0.44 _{-0.10} ^{+0.18}$  &  $ 1.10 _{-0.22} ^{+0.23}$  &  $  0.045 _{-0.008} ^{+0.014}$ \\
0923$+$3028  &  $0.04495 \pm 0.00049$  &  $296.0 \pm 3.0$  &  $0.275 \pm 0.015$ & $ 0.36 _{-0.09} ^{+0.14}$  &  $ 1.04 _{-0.22} ^{+0.24}$  &  $  0.059 _{-0.011} ^{+0.017}$ \\
1436$+$5010  &  $0.04580 \pm 0.00010$  &  $347.4 \pm 8.9$  &  $0.234 \pm 0.013$ & $ 0.30 _{-0.07} ^{+0.10}$  &  $ 1.02 _{-0.20} ^{+0.23}$  &  $  0.071 _{-0.012} ^{+0.017}$ \\
0825$+$1152  &  $0.05819 \pm 0.00001$  &  $319.4 \pm 2.7$  &  $0.279 \pm 0.021$ & $ 0.35 _{-0.08} ^{+0.11}$  &  $ 1.07 _{-0.19} ^{+0.23}$  &  $  0.113 _{-0.019} ^{+0.026}$ \\
1741$+$6526  &  $0.06111 \pm 0.00001$  &  $508.0 \pm 4.0$  &  $0.170 \pm 0.010$ & $ 0.14 _{-0.01} ^{+0.00}$  &  $ 1.34 _{-0.04} ^{+0.07}$  &  $  0.154 _{-0.006} ^{+0.004}$ \\
0755$+$4906  &  $0.06302 \pm 0.00213$  &  $438.0 \pm 5.0$  &  $0.184 \pm 0.010$ & $ 0.19 _{-0.03} ^{+0.02}$  &  $ 1.15 _{-0.10} ^{+0.17}$  &  $  0.178 _{-0.019} ^{+0.015}$ \\
2338$-$2052  &  $0.07644 \pm 0.00712$  &  $133.4 \pm 7.5$  &  $0.258 \pm 0.016$ & $ 0.34 _{-0.09} ^{+0.16}$  &  $ 1.01 _{-0.24} ^{+0.25}$  &  $  0.261 _{-0.049} ^{+0.089}$ \\
2309$+$2603  &  $0.07653 \pm 0.00001$  &  $405.8 \pm 3.5$  &  $0.176 \pm 0.010$ & $ 0.19 _{-0.03} ^{+0.02}$  &  $ 1.11 _{-0.10} ^{+0.17}$  &  $  0.317 _{-0.036} ^{+0.029}$ \\
0849$+$0445  &  $0.07870 \pm 0.00010$  &  $366.9 \pm 4.7$  &  $0.179 \pm 0.010$ & $ 0.21 _{-0.04} ^{+0.04}$  &  $ 1.04 _{-0.14} ^{+0.19}$  &  $  0.359 _{-0.048} ^{+0.049}$ \\
0751$-$0141  &  $0.08001 \pm 0.00279$  &  $432.6 \pm 2.3$  &  $0.194 \pm 0.010$ & $ 0.20 _{-0.00} ^{+0.00}$  &  $ 1.17 _{-0.01} ^{+0.01}$  &  $  0.317 _{-0.004} ^{+0.003}$ \\
2119$-$0018  &  $0.08677 \pm 0.00004$  &  $383.0 \pm 4.0$  &  $0.159 \pm 0.010$ & $ 0.19 _{-0.03} ^{+0.01}$  &  $ 0.99 _{-0.05} ^{+0.14}$  &  $  0.530 _{-0.053} ^{+0.024}$ \\
1234$-$0228  &  $0.09143 \pm 0.00400$  &  $ 94.0 \pm 2.3$  &  $0.227 \pm 0.014$ & $ 0.30 _{-0.07} ^{+0.14}$  &  $ 0.97 _{-0.24} ^{+0.24}$  &  $  0.478 _{-0.091} ^{+0.160}$ \\
1054$-$2121  &  $0.10439 \pm 0.00655$  &  $261.1 \pm 7.1$  &  $0.178 \pm 0.011$ & $ 0.23 _{-0.05} ^{+0.09}$  &  $ 0.95 _{-0.21} ^{+0.23}$  &  $  0.829 _{-0.141} ^{+0.217}$ \\
0745$+$1949  &  $0.11240 \pm 0.00833$  &  $108.7 \pm 2.9$  &  $0.164 \pm 0.010$ & $ 1.11 _{-0.77} ^{+0.27}$  &  $ 0.31 _{-0.03} ^{+0.33}$  &  $  3.95  _{-2.41}  ^{+0.81} $ \\
1108$+$1512  &  $0.12310 \pm 0.00867$  &  $256.2 \pm 3.7$  &  $0.179 \pm 0.010$ & $ 0.23 _{-0.05} ^{+0.08}$  &  $ 0.96 _{-0.21} ^{+0.22}$  &  $  1.27  _{-0.21}  ^{+0.32} $ \\
0112$+$1835  &  $0.14698 \pm 0.00003$  &  $295.3 \pm 2.0$  &  $0.160 \pm 0.010$ & $ 0.22 _{-0.04} ^{+0.02}$  &  $ 0.90 _{-0.05} ^{+0.15}$  &  $  2.35  _{-0.29}  ^{+0.13} $ \\
1233$+$1602  &  $0.15090 \pm 0.00009$  &  $336.0 \pm 4.0$  &  $0.169 \pm 0.010$ & $ 0.17 _{-0.02} ^{+0.02}$  &  $ 1.15 _{-0.09} ^{+0.16}$  &  $  1.96  _{-0.20}  ^{+0.15} $ \\
1130$+$3855  &  $0.15652 \pm 0.00001$  &  $284.0 \pm 4.9$  &  $0.288 \pm 0.018$ & $ 0.32 _{-0.06} ^{+0.05}$  &  $ 1.18 _{-0.12} ^{+0.19}$  &  $  1.40  _{-0.19}  ^{+0.18} $ \\
1112$+$1117  &  $0.17248 \pm 0.00001$  &  $116.2 \pm 2.8$  &  $0.176 \pm 0.010$ & $ 0.23 _{-0.06} ^{+0.11}$  &  $ 0.93 _{-0.24} ^{+0.24}$  &  $  3.25  _{-0.59}  ^{+1.06} $ \\
1005$+$3550  &  $0.17652 \pm 0.00011$  &  $143.0 \pm 2.3$  &  $0.168 \pm 0.010$ & $ 0.22 _{-0.05} ^{+0.10}$  &  $ 0.92 _{-0.23} ^{+0.24}$  &  $  3.62  _{-0.66}  ^{+1.15} $ \\
0818$+$3536  &  $0.18315 \pm 0.02110$  &  $170.0 \pm 5.0$  &  $0.165 \pm 0.010$ & $ 0.22 _{-0.05} ^{+0.10}$  &  $ 0.92 _{-0.23} ^{+0.24}$  &  $  4.05  _{-0.73}  ^{+1.26} $ \\
1443$+$1509  &  $0.19053 \pm 0.02402$  &  $306.7 \pm 3.0$  &  $0.201 \pm 0.013$ & $ 0.20 _{-0.03} ^{+0.02}$  &  $ 1.20 _{-0.09} ^{+0.16}$  &  $  3.06  _{-0.31}  ^{+0.27} $ \\
1840$+$6423  &  $0.19130 \pm 0.00005$  &  $272.0 \pm 2.0$  &  $0.182 \pm 0.011$ & $ 0.21 _{-0.04} ^{+0.04}$  &  $ 1.04 _{-0.14} ^{+0.20}$  &  $  3.76  _{-0.52}  ^{+0.52} $ \\
2103$-$0027  &  $0.20308 \pm 0.00023$  &  $281.0 \pm 3.2$  &  $0.161 \pm 0.010$ & $ 0.18 _{-0.03} ^{+0.03}$  &  $ 1.05 _{-0.13} ^{+0.19}$  &  $  4.85  _{-0.61}  ^{+0.57} $ \\
1238$+$1946  &  $0.22275 \pm 0.00009$  &  $258.6 \pm 2.5$  &  $0.210 \pm 0.011$ & $ 0.24 _{-0.04} ^{+0.04}$  &  $ 1.08 _{-0.13} ^{+0.19}$  &  $  4.91  _{-0.66}  ^{+0.62} $ \\
1249$+$2626  &  $0.22906 \pm 0.00112$  &  $191.6 \pm 3.9$  &  $0.160 \pm 0.010$ & $ 0.21 _{-0.05} ^{+0.08}$  &  $ 0.92 _{-0.22} ^{+0.23}$  &  $  7.51  _{-1.31}  ^{+2.09} $ \\
0345$+$1748  &  $0.23503 \pm 0.00013$  &  $273.4 \pm 0.5$  &  $0.218 \pm 0.012$ & $ 0.27 _{-0.01} ^{+0.01}$  &  $ 1.02 _{-0.02} ^{+0.02}$  &  $  5.80  _{-0.21}  ^{+0.23} $ \\
0822$+$2753  &  $0.24400 \pm 0.00020$  &  $271.1 \pm 9.0$  &  $0.191 \pm 0.012$ & $ 0.21 _{-0.03} ^{+0.03}$  &  $ 1.12 _{-0.11} ^{+0.17}$  &  $  6.52  _{-0.77}  ^{+0.63} $ \\
1631$+$0605  &  $0.24776 \pm 0.00411$  &  $215.4 \pm 3.4$  &  $0.162 \pm 0.010$ & $ 0.21 _{-0.05} ^{+0.07}$  &  $ 0.95 _{-0.19} ^{+0.22}$  &  $  8.94  _{-1.46}  ^{+2.05} $ \\
1526$+$0543  &  $0.25039 \pm 0.00001$  &  $231.9 \pm 2.3$  &  $0.161 \pm 0.010$ & $ 0.20 _{-0.04} ^{+0.05}$  &  $ 0.97 _{-0.17} ^{+0.22}$  &  $  9.04  _{-1.41}  ^{+1.72} $ \\
2132$+$0754  &  $0.25056 \pm 0.00002$  &  $297.3 \pm 3.0$  &  $0.187 \pm 0.010$ & $ 0.17 _{-0.02} ^{+0.01}$  &  $ 1.26 _{-0.07} ^{+0.13}$  &  $  6.43  _{-0.52}  ^{+0.33} $ \\
1141$+$3850  &  $0.25958 \pm 0.00005$  &  $265.8 \pm 3.5$  &  $0.177 \pm 0.010$ & $ 0.19 _{-0.03} ^{+0.03}$  &  $ 1.10 _{-0.11} ^{+0.18}$  &  $  8.29  _{-0.97}  ^{+0.82} $ \\
1630$+$2712  &  $0.27646 \pm 0.00002$  &  $218.0 \pm 5.0$  &  $0.170 \pm 0.010$ & $ 0.21 _{-0.05} ^{+0.06}$  &  $ 0.97 _{-0.17} ^{+0.22}$  &  $ 11.3   _{-1.8}   ^{+2.2}  $ \\
1449$+$1717  &  $0.29075 \pm 0.00001$  &  $228.5 \pm 3.2$  &  $0.171 \pm 0.010$ & $ 0.21 _{-0.04} ^{+0.05}$  &  $ 1.00 _{-0.15} ^{+0.20}$  &  $ 12.5   _{-1.8}   ^{+2.0}  $ \\
0917$+$4638  &  $0.31642 \pm 0.00002$  &  $148.8 \pm 2.0$  &  $0.173 \pm 0.010$ & $ 0.23 _{-0.05} ^{+0.10}$  &  $ 0.94 _{-0.23} ^{+0.23}$  &  $ 16.5   _{-2.8}   ^{+5.0}  $ \\
0152$+$0749  &  $0.32288 \pm 0.00014$  &  $217.0 \pm 2.0$  &  $0.169 \pm 0.010$ & $ 0.20 _{-0.04} ^{+0.05}$  &  $ 1.00 _{-0.16} ^{+0.21}$  &  $ 16.8   _{-2.5}   ^{+3.0}  $ \\
1422$+$4352  &  $0.37930 \pm 0.01123$  &  $176.0 \pm 6.0$  &  $0.181 \pm 0.010$ & $ 0.23 _{-0.05} ^{+0.08}$  &  $ 0.96 _{-0.21} ^{+0.22}$  &  $ 25.2   _{-4.3}   ^{+6.4}  $ \\
1617$+$1310  &  $0.41124 \pm 0.00086$  &  $210.1 \pm 2.8$  &  $0.172 \pm 0.010$ & $ 0.20 _{-0.04} ^{+0.04}$  &  $ 1.02 _{-0.14} ^{+0.20}$  &  $ 30.8   _{-4.3}   ^{+4.4}  $ \\
1538$+$0252  &  $0.41915 \pm 0.00295$  &  $227.6 \pm 4.9$  &  $0.168 \pm 0.010$ & $ 0.18 _{-0.03} ^{+0.03}$  &  $ 1.09 _{-0.11} ^{+0.18}$  &  $ 31.4   _{-3.7}   ^{+3.1}  $ \\
1439$+$1002  &  $0.43741 \pm 0.00169$  &  $174.0 \pm 2.0$  &  $0.181 \pm 0.010$ & $ 0.23 _{-0.05} ^{+0.08}$  &  $ 0.97 _{-0.19} ^{+0.22}$  &  $ 36.8   _{-6.0}   ^{+8.6}  $ \\
0837$+$6648  &  $0.46329 \pm 0.00005$  &  $150.3 \pm 3.0$  &  $0.181 \pm 0.010$ & $ 0.24 _{-0.06} ^{+0.09}$  &  $ 0.95 _{-0.22} ^{+0.23}$  &  $ 43.7   _{-7.7}  ^{+12.0} $ \\
0940$+$6304  &  $0.48438 \pm 0.00001$  &  $210.4 \pm 3.2$  &  $0.180 \pm 0.010$ & $ 0.20 _{-0.03} ^{+0.03}$  &  $ 1.08 _{-0.12} ^{+0.18}$  &  $ 43.9   _{-5.4}   ^{+4.8}  $ \\
0840$+$1527  &  $0.52155 \pm 0.00474$  &  $ 84.8 \pm 3.1$  &  $0.192 \pm 0.010$ & $ 0.25 _{-0.06} ^{+0.12}$  &  $ 0.95 _{-0.24} ^{+0.24}$  &  $ 57.5   _{-10.7}  ^{+18.2} $ \\
0802$-$0955  &  $0.54687 \pm 0.00455$  &  $176.5 \pm 4.5$  &  $0.198 \pm 0.012$ & $ 0.24 _{-0.05} ^{+0.06}$  &  $ 1.02 _{-0.17} ^{+0.22}$  &  $ 59.3   _{-9.1}  ^{+11.0} $ \\
1518$+$1354  &  $0.57655 \pm 0.00734$  &  $112.7 \pm 4.6$  &  $0.147 \pm 0.018$ & $ 0.19 _{-0.05} ^{+0.09}$  &  $ 0.90 _{-0.24} ^{+0.26}$  &  $ 97.4   _{-19.4}  ^{+31.0} $ \\
2151$+$1614  &  $0.59152 \pm 0.00008$  &  $163.3 \pm 3.1$  &  $0.181 \pm 0.010$ & $ 0.23 _{-0.05} ^{+0.07}$  &  $ 0.97 _{-0.18} ^{+0.22}$  &  $ 81.7   _{-13.3}  ^{+17.3} $ \\
1512$+$2615  &  $0.59999 \pm 0.02348$  &  $115.0 \pm 4.0$  &  $0.250 \pm 0.014$ & $ 0.33 _{-0.08} ^{+0.14}$  &  $ 1.01 _{-0.22} ^{+0.24}$  &  $ 65.1   _{-12.1}  ^{+19.7} $ \\
1518$+$0658  &  $0.60935 \pm 0.00004$  &  $172.0 \pm 2.0$  &  $0.224 \pm 0.013$ & $ 0.27 _{-0.05} ^{+0.06}$  &  $ 1.06 _{-0.15} ^{+0.21}$  &  $ 70.0   _{-10.4}  ^{+11.4} $ \\
0756$+$6704  &  $0.61781 \pm 0.00002$  &  $204.2 \pm 1.6$  &  $0.182 \pm 0.011$ & $ 0.19 _{-0.03} ^{+0.02}$  &  $ 1.14 _{-0.10} ^{+0.17}$  &  $ 79.8   _{-8.7}   ^{+7.1}  $ \\
1151$+$5858  &  $0.66902 \pm 0.00070$  &  $175.7 \pm 5.9$  &  $0.186 \pm 0.011$ & $ 0.22 _{-0.04} ^{+0.04}$  &  $ 1.03 _{-0.14} ^{+0.19}$  &  $105     _{-15}    ^{+15}   $ \\
0730$+$1703  &  $0.69770 \pm 0.05427$  &  $122.8 \pm 4.3$  &  $0.182 \pm 0.010$ & $ 0.24 _{-0.06} ^{+0.10}$  &  $ 0.94 _{-0.22} ^{+0.23}$  &  $130     _{-23}    ^{+38}   $ \\
0308$+$5140  &  $0.80590 \pm 0.00109$  &  $ 78.9 \pm 2.7$  &  $0.149 \pm 0.015$ & $ 0.20 _{-0.05} ^{+0.09}$  &  $ 0.90 _{-0.24} ^{+0.25}$  &  $233     _{-45}    ^{+75}   $ \\
0811$+$0225  &  $0.82194 \pm 0.00049$  &  $220.7 \pm 2.5$  &  $0.179 \pm 0.010$ & $ 0.14 _{-0.01} ^{+0.01}$  &  $ 1.45 _{-0.05} ^{+0.10}$  &  $141     _{-7}    ^{+4}   $ \\
1241$+$0633  &  $0.95912 \pm 0.00028$  &  $138.2 \pm 4.8$  &  $0.199 \pm 0.012$ & $ 0.25 _{-0.05} ^{+0.07}$  &  $ 1.00 _{-0.18} ^{+0.22}$  &  $270     _{-44}    ^{+58}   $ \\
2236$+$2232  &  $1.01016 \pm 0.00005$  &  $119.9 \pm 2.0$  &  $0.186 \pm 0.010$ & $ 0.24 _{-0.06} ^{+0.09}$  &  $ 0.96 _{-0.21} ^{+0.23}$  &  $338     _{-58}    ^{+91}   $ \\
0815$+$2309  &  $1.07357 \pm 0.00018$  &  $131.7 \pm 2.6$  &  $0.200 \pm 0.021$ & $ 0.25 _{-0.06} ^{+0.08}$  &  $ 1.00 _{-0.19} ^{+0.24}$  &  $367     _{-66}    ^{+85}   $ \\
\enddata
\end{deluxetable*}

	\clearpage

\appendix \section{DATA TABLES}

	Table \ref{tab:dat} presents the radial velocity measurements for our 15 new 
ELM WD candidates.  Table \ref{tab:dat} columns include object name, heliocentric 
Julian date (based on UTC), and heliocentric radial velocity (uncorrected for the WD 
gravitational redshift).

	Table \ref{tab:elmsurvey} presents the entire ELM Survey sample of 88 
objects.  Table \ref{tab:elmsurvey} columns include object name, \teff, \logg, 
estimated ELM WD mass and absolute magnitude $M_g$, de-reddened apparent magnitude 
$g_0$, heliocentric distance $d_{helio}$, orbital period $P$, semi-amplitude $k$, 
systemic velocity $\gamma$, minimum secondary mass $M_2$, and maximum gravitational 
wave merger time $\tau$.

\begin{deluxetable}{ccc}	
\tablecolumns{3}
\tablewidth{0pt}
\tablecaption{Radial Velocity Measurements\label{tab:dat}}
\tablehead{
	\colhead{Object}& \colhead{HJD}& \colhead{$v_{helio}$}\\
			& +2450000     & (\kms )
}
	\startdata
J0125+2017 & 6329.595091 & $   30.9 \pm  7.7 $ \\
\nodata    & 6329.612901 & $   15.8 \pm  6.0 $ \\
\nodata    & 6330.630689 & $   41.9 \pm  5.0 $ \\
\nodata    & 6331.595507 & $   73.8 \pm  3.9 $ \\
\nodata    & 7008.555545 & $    0.1 \pm  7.9 $ \\
	\enddata

	\tablecomments{This table is available in its entirety in 
machine-readable and Virtual Observatory forms in the online journal. A portion is 
shown here for guidance regarding its form and content.}

\end{deluxetable}

\begin{deluxetable}{cccccccccccccc}	
\tabletypesize{\scriptsize}
\tablecolumns{14}
\tablewidth{0pt}
\tablecaption{Entire ELM Survey\label{tab:elmsurvey}}
\tablehead{
	\colhead{Object} &
	\colhead{R.A.} &
	\colhead{Decl.} &
	\colhead{$T_{\rm eff}$} &
	\colhead{$\log g$} &
	\colhead{Mass} &
	\colhead{$M_g$} &
	\colhead{$g_0$} &
	\colhead{$d_{helio}$} &
	\colhead{$P$} &
	\colhead{$k$} &
	\colhead{$\gamma$} &
	\colhead{$M_2$} &
	\colhead{$\tau$}\\
 \colhead{} & \colhead{(h:m:s)} & \colhead{(d:m:s)} & \colhead{(K)} & \colhead{(cm s$^{-2}$)} & \colhead{(\msun)} & \colhead{(mag)} & \colhead{(mag)} & \colhead{(kpc)}
  & \colhead{(days)} & \colhead{(km s$^{-1}$)} & \colhead{(km s$^{-1}$)} & \colhead{(\msun)} & \colhead{(Gyr)}
}
\startdata
J0022$+$0031 &  0:22:28.452 &   0:31:15.55 & $20460 \pm 310$ & $7.58 \pm 0.04$ & 0.459 & $ 9.88 \pm 0.08$ & $19.284 \pm 0.034$ & $0.762 \pm 0.029$ & $0.49135 \pm 0.02540$ & $ 80.8 \pm  1.3$ & $ -20.3 \pm  0.8$ & $>$0.23 & $<$59.25 \\
J0022$-$1014 &  0:22:07.659 & -10:14:23.53 & $20730 \pm 340$ & $7.28 \pm 0.05$ & 0.376 & $ 9.32 \pm 0.10$ & $19.581 \pm 0.031$ & $1.129 \pm 0.052$ & $0.07989 \pm 0.00300$ & $145.6 \pm  5.6$ & $ -38.5 \pm  3.7$ & $>$0.21 & $<$0.613 \\
J0056$-$0611 &  0:56:48.232 &  -6:11:41.59 & $12230 \pm 180$ & $6.17 \pm 0.04$ & 0.180 & $ 8.37 \pm 0.10$ & $17.208 \pm 0.023$ & $0.586 \pm 0.029$ & $0.04338 \pm 0.00002$ & $376.9 \pm  2.4$ & $   4.2 \pm  1.2$ & $>$0.46 & $<$0.115 \\
J0106$-$1000 &  1:06:57.398 & -10:00:03.35 & $16970 \pm 260$ & $6.10 \pm 0.05$ & 0.189 & $ 7.47 \pm 0.14$ & $19.595 \pm 0.023$ & $2.669 \pm 0.176$ & $0.02715 \pm 0.00002$ & $395.2 \pm  3.6$ & $   2.2 \pm  2.7$ & $>$0.39 & $<$0.036 \\
J0112$+$1835 &  1:12:10.254 &  18:35:03.77 & $10020 \pm 140$ & $5.76 \pm 0.05$ & 0.160 & $ 8.00 \pm 0.12$ & $17.110 \pm 0.016$ & $0.664 \pm 0.036$ & $0.14698 \pm 0.00003$ & $295.3 \pm  2.0$ & $-121.4 \pm  1.1$ & $>$0.62 & $<$2.674 \\
\enddata
\tablecomments{Table 5 is published in its entirety in the electronic 
edition of the {\it Astrophysical Journal}.  A portion is shown here 
for guidance regarding its form and content.}
\end{deluxetable}

\end{document}